\documentclass[10pt,english,floatfix,superscriptaddress,aps,prd,preprint,showkeys,nofootinbib]{revtex4}
\usepackage{amsmath}
\usepackage{amssymb}
\usepackage{amsbsy}
\usepackage{amsfonts}
\usepackage{amsopn}
\usepackage{amstext}
\usepackage{graphicx}
\usepackage[english]{babel}
\usepackage{color}
\usepackage{slashed}
\usepackage{esint}
\usepackage[dvips]{epsfig}
\usepackage[dvips]{graphicx}
\usepackage{float}
\usepackage{units}
\usepackage{textcomp}
\usepackage{wasysym}
\usepackage{hyperref}
\usepackage{slashed}

\begin{document}

\title{Hairy Kiselev black hole with quintessential matter: themodynamic properties, sparsity of Hawking radiation, and greybody factors}

\author{Faizuddin Ahmed}
\email{faizuddinahmed15@gmail.com}
\affiliation{Department of Physics, The Assam Royal Global University, 781035, Guwahati, Assam, India}

\author{Fernando M. Belchior}
\email{fernandobelcks7@gmail.com}
\affiliation{Departamento de Física, Universidade Federal da Paraíba, Centro de Ciências Exatas e da Natureza, 58051-970, João Pessoa, Paraíba, Brazil}

\author{Allan R. P. Moreira}
\email{allan.moreira@fisica.ufc.br}
\affiliation{Secretaria da Educação do Ceará (SEDUC), Coordenadoria Regional de Desenvolvimento da Educação (CREDE
9), 62880-384, Horizonte, Ceará, Brazil}

\author{Abdelmalek Bouzenada}
\email{abdelmalekbouzenada@gmail.com}
\affiliation{Laboratory of Theoretical and Applied Physics, Echahid Cheikh Larbi Tebessi University, 12001, Algeria}

\begin{abstract}
In this work, we investigate the thermodynamic properties, Hawking-radiation sparsity, and greybody-factor bounds of a hairy Kiselev black hole surrounded by a quintessential fluid. The spacetime geometry contains, in addition to the Schwarzschild mass term, a quintessence contribution controlled by the intensity parameter $N$ and the state parameter $\omega_q$, an exponential hair correction governed by the coupling $\alpha$ and the primary hair scale $\ell$, and a cosmological constant $\Lambda$. We first examined the horizon structure and derived the black hole mass from the horizon condition. The Hawking temperature, heat capacity, and Gibbs free energy were then analyzed in order to identify the influence of the hair and surrounding-field parameters on local and global thermodynamic behavior. The results show that the exponential hair mainly affects the small-horizon regime, while the quintessential sector and the cosmological constant produce significant changes in the large-scale behavior of the black hole. In particular, the heat capacity exhibits divergences associated with second-order phase transitions, and the Gibbs free energy reveals the possibility of competing thermodynamic branches. We also analyzed the sparsity of Hawking radiation and showed that the emitted flux is highly intermittent rather than continuous. The sparsity parameter is controlled by the combination of the Hawking temperature, the effective emitting area, and the greybody factors. Finally, we studied massless scalar perturbations in this background by reducing the Klein-Gordon equation to a Schrödinger-like radial equation with an effective potential. This allowed us to discuss the stability of the spacetime and to obtain rigorous lower bounds on the greybody factor. The analysis shows that higher angular modes are more strongly suppressed by the centrifugal barrier, whereas the quintessence and hair parameters modify the transmission probability through their effect on the height and width of the effective potential.
\end{abstract}


\maketitle

\section{Introduction} \label{Sec1}
The most recent observational breakthroughs have provided compelling and direct evidence for the existence of black holes (BHs), transforming them from theoretical constructs into astrophysical realities \cite{EventHorizonTelescope:2019dse,EventHorizonTelescope:2019ggy}. Over a century has passed since the Schwarzschild solution to Einstein’s field equations (EFEs) first predicted the existence of these exotic objects within the framework of General Relativity (GR) \cite{Schwarzschild:1916uq}. Since then, BHs have evolved from mere mathematical solutions to fundamental components of our understanding of the universe. Beyond the standard solutions of GR, a wide array of BH models have been proposed within the context of modified and alternative theories of gravity. These include, but are not limited to, scalar-tensor theories, Einstein-Gauss-Bonnet (EGB) gravity, f(R) gravity, and massive gravity frameworks \cite{Bronnikov:1973fh,Maartens:2003tw,DeFelice:2010aj,Harko:2011kv,Capozziello:2011et,Sotiriou:2013qea}. These alternative models aim to address the limitations of GR, such as the nature of dark energy, singularity resolution, or quantum gravity effects \cite{Donoghue:1994dn,Satheeshkumar:2021zvl,Carballo-Rubio:2018jzw,Errahmani:2024ran}.

The first quantitative prediction of light deflection by a gravitational field was made by Johann Georg von Soldner in 1801 within the framework of Newtonian mechanics \cite{soldner1804deflection}. By treating light as a stream of particles, he estimated that a light ray passing near the Sun would experience a deflection of approximately 0.87 arc seconds. Over a century later, Albert Einstein revisited the problem using the principles of GR, which accounts for the curvature of both space and time. His calculations yielded a deflection angle of 1.75 arc seconds, precisely twice the Newtonian prediction. This theoretical advancement was empirically validated during the 1919 total solar eclipse by Sir Arthur Eddington and his team, who measured the apparent shift in the positions of stars near the Sun's limb \cite{Dyson:1920cwa}. Their observations closely matched Einstein’s prediction, providing the first direct experimental confirmation of GR and marking a pivotal moment in the history of modern physics. The earliest analytical investigation of light deflection in the strong-field regime, particularly near the photon sphere, was conducted by Darwin in 1959 \cite{darwin1959gravity}, where he obtained an approximate expression for the bending angle by expanding around the unstable circular photon orbit. Following Darwin’s foundational work, several researchers advanced the theoretical understanding of gravitational light bending in the strong-field regime. Atkinson \cite{atkinson1965light} investigated light propagation near compact objects and explored photon trajectories under extreme gravitational fields. Luminet \cite{luminet1979image} provided one of the first detailed simulations of BH shadows and the visual appearance of accretion disks, incorporating relativistic lensing effects. Chandrasekhar \cite{chandrasekhar1998mathematical}, in his seminal monograph, presented a comprehensive mathematical treatment of photon orbits and deflection angles in Schwarzschild and Kerr geometries. Ohanian \cite{ohanian1987black} refined earlier estimates by offering improved analytical expressions for the deflection angle and clarified the physical interpretation of the bending angle in curved spacetimes.

In the early 2000s, a resurgence of interest in strong gravitational lensing led to significant theoretical advancements. Virbhadra and Ellis \cite{virbhadra2000schwarzschild} introduced a lensing framework for strong deflection by BHs and demonstrated that relativistic images could be formed near the photon sphere. Concurrently, Frittelli, Kling, and Newman \cite{frittelli2000spacetime} provided a rigorous mathematical foundation by deriving lens equations directly from spacetime geometry using the null geodesic structure. Eiroa, Romero, and Torres \cite{eiroa2002reissner}  extended these studies by analyzing strong lensing observables for various BH metrics. Petters \cite{petters2003relativistic} contributed to the mathematical formalism by developing a general theory for gravitational lensing in the strong-field limit using singularity theory. Perlick \cite{perlick2004exact} focused on exact lens equations and classification of caustics in spherically symmetric and static spacetimes. Bozza and collaborators \cite{bozza2001strong} made extensive contributions by formulating the strong deflection limit (SDL) approach, which provides analytical expressions for deflection angles and observable parameters in various BH backgrounds. In addition, S.V. Iyer and A.O. Petters  applied the SDL formalism to both Schwarzschild and Kerr BHs, deriving explicit expressions for deflection angles and lensing coefficients, thus highlighting spin-induced asymmetry in Kerr lensing \cite{Iyer:2006cn,Iyer:2009wa}. Uniyal et al. \cite{Uniyal:2018ngj} extended this formalism to the Kerr–Sen BH, exploring the effect of dilaton-axion fields on gravitational lensing and providing a deeper understanding of string-theoretic corrections. More recently, Y.W. Hsiao et al. \cite{Hsiao:2019ohy} investigated gravitational lensing by Kerr–Newman BHs, offering a comprehensive treatment of deflection angles in both weak and strong field limits, and analyzing how charge and spin jointly influence image formation. Numerous studies have been conducted on gravitational lensing, among which a few representative works have been cited herein \cite{Ishihara:2016vdc,Beloborodov:2002mr,Bjerrum-Bohr:2014zsa,Ovgun:2019wej,Cunha:2019hzj,Shaikh:2019itn,Sotani:2015ewa,Moffat:2008gi,Liu:2015zou,Majumdar:2004mz,Tsukamoto:2014dta,Fu:2021fxn,Heydari-Fard:2021pjc,Kumaran:2023brp,Paul:2020ufc,Lim:2016lqv,Kala:2020prt,Kala:2020viz,Parbin:2023zik,Javed:2023iih,Nazari:2022fbn,Kala:2022uog,Ovgun:2019qzc,Ovgun:2025ctx,Pantig:2022ely,Kuang:2022xjp,Javed_2020,Li:2020dln,Abdujabbarov:2017pfw,Atamurotov:2023rye,Kala:2024fvg,Wang:2024iwt,Islam:2022ybr,Kumar:2020hgm,Kumar:2020sag,Vishvakarma:2024icz,Pantig:2024ixc,Kala:2025iri,Roy:2025hdw}.

In modern cosmology, the accelerated expansion of the Universe---first confirmed through observations of \textit{Type Ia supernovae}---has been attributed to a mysterious component known as \textit{dark energy}. Within the standard $\Lambda$CDM model, this is typically modeled by a cosmological constant ($\Lambda$). However, dynamical alternatives such as \textit{quintessence}, a slowly rolling scalar field with an evolving equation of state $w_q \in (-1, -1/3)$, have been proposed to address the fine-tuning and coincidence problems associated with $\Lambda$. When BHs are considered in such non-vacuum backgrounds, particularly surrounded by quintessence-like fields, the spacetime geometry is significantly altered from classical solutions like Schwarzschild or Kerr. These modifications have direct consequences on the propagation of photons, leading to measurable changes in \textit{gravitational lensing observables}, such as the bending angle, photon sphere structure, and the location of relativistic images.

Gravitational lensing by BHs in a quintessence-dominated background therefore provides a compelling framework to study deviations from general relativity and standard cosmology. This becomes particularly relevant in the era of precision observations through instruments like the \textit{Event Horizon Telescope (EHT)}, \textit{Square Kilometre Array (SKA)}, and future space-based missions like \textit{LISA}, which are capable of probing the near-horizon structure of compact objects. Accurate modeling of such systems may reveal subtle imprints of dark energy fields, providing complementary constraints to cosmological measurements from supernovae and the cosmic microwave background (CMB). In this context, analyzing light deflection in BH spacetimes influenced by quintessence not only enriches our understanding of strong-field gravity but also connects astrophysical observations with the underlying dynamics of the Universe’s expansion. Gravitational lensing around BHs in the presence of quintessence has garnered significant interest, and various authors have contributed to this area. Among these, we have cited several relevant works\cite{Younas_2015,ABBAS2021100750,Javed_2020,doi:10.1142/S0217732321502242,kala2025nullgeodesicsthermodynamicsweak,Azreg_A_nou_2017,Mustafa_2022, Finelli:2006iz,Liu:2008sg,Fernando:2014rsa,Atamurotov:2022knb,Molla:2023yxn}.

\section{Hairy Kiselev AdS BH Spacetime} \label{Sec2}

V.V.Kiselev proposed an exact static spherically symmetric solution \cite{Kiselev_2003} to Einstein's field equations, incorporating the state parameter $\omega$. Using the method of extended gravitational decoupling \cite{PhysRevD.95.104019,OVALLE2019213,OVALLE2021100744}, a modified metric was introduced with extra hairs. The metric is given by\cite{PhysRevD.108.044073},
\begin{equation}
    ds^2 = -f(r)\,dt^2 + \frac{1}{f(r)}\,dr^2 + r^2\,d\Omega^2,\label{metric}
\end{equation}
where, the metric function is given by
\begin{equation}
    f(r) = 1- \frac{2\,M}{r}  - \frac{N}{r^{3\,\omega + 1}} + \alpha\, \exp\left(-\frac{r}{M- \frac{\alpha\,\ell}{2}}\right)-\frac{\Lambda}{3}\,r^2,\label{function}
\end{equation}
and \[
d\Omega^2 = d\theta^2 + \sin^2\theta\ d\phi^2.
\]
Here, $M$ is the mass of the BH and is related to the Schwarzschild mass $\mathcal{M}$ as $M = \mathcal{M}+\frac{\alpha\, \ell}{2}$,\, $\alpha$ is the coupling constant, $\ell$ is a constant parameter with length dimensions associated with the primary hair of the BH, $\omega $ is the state parameter which takes different values depending upon the nature of the surrounding field and $N$ is a constant related to the strength of the surrounding field.  $\omega$ and $N$ must have different signs in order to respect the weak energy conditions.

\begin{figure}[ht!]
    \centering
    \includegraphics[height=5cm]{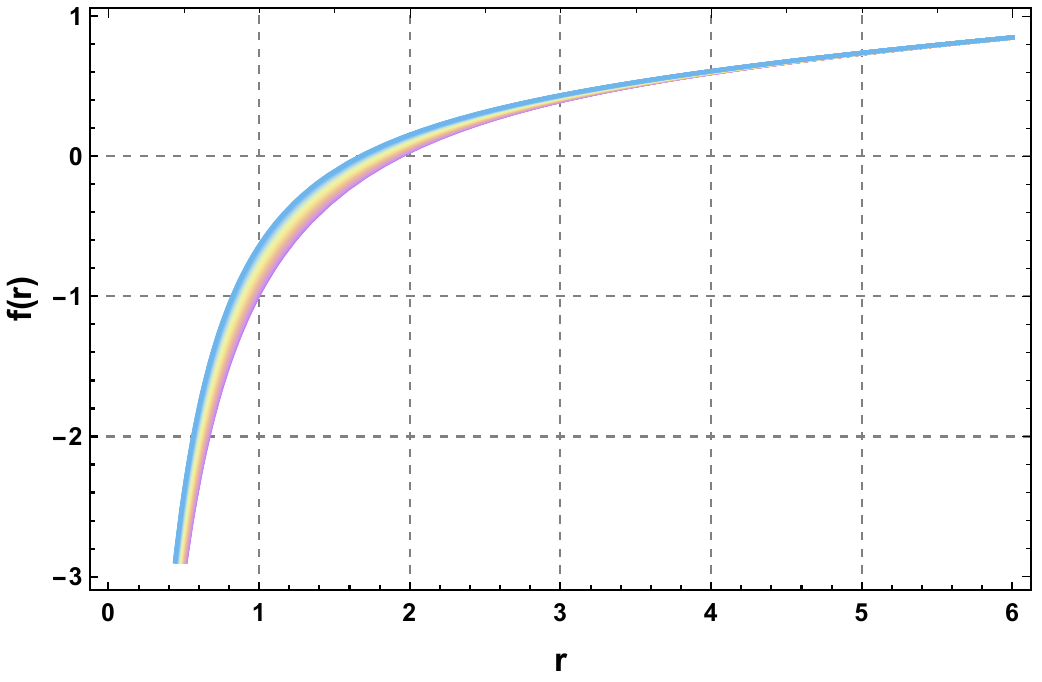}
    \includegraphics[height=5cm]{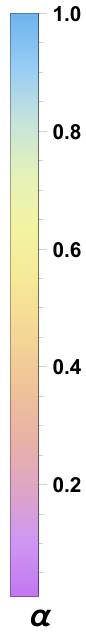}\qquad
    \includegraphics[height=5cm]{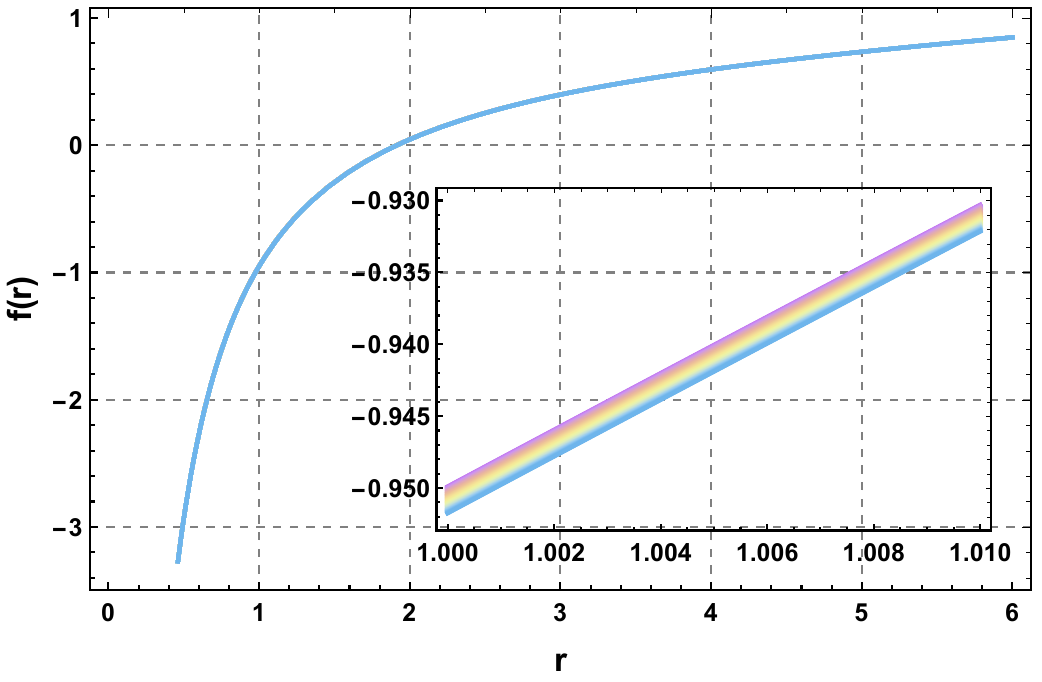}
    \includegraphics[height=5cm]{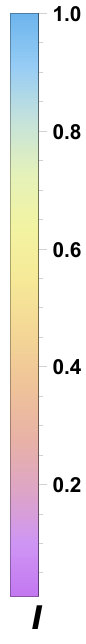}\\
    (a) $\ell=0.1$, $\omega=-2/3$ and  $N=-\Lambda=0.01$ \hspace{3cm} (b) $\alpha=0.1$, $\omega=-2/3$ and  $N=-\Lambda=0.01$\\
    \includegraphics[height=5cm]{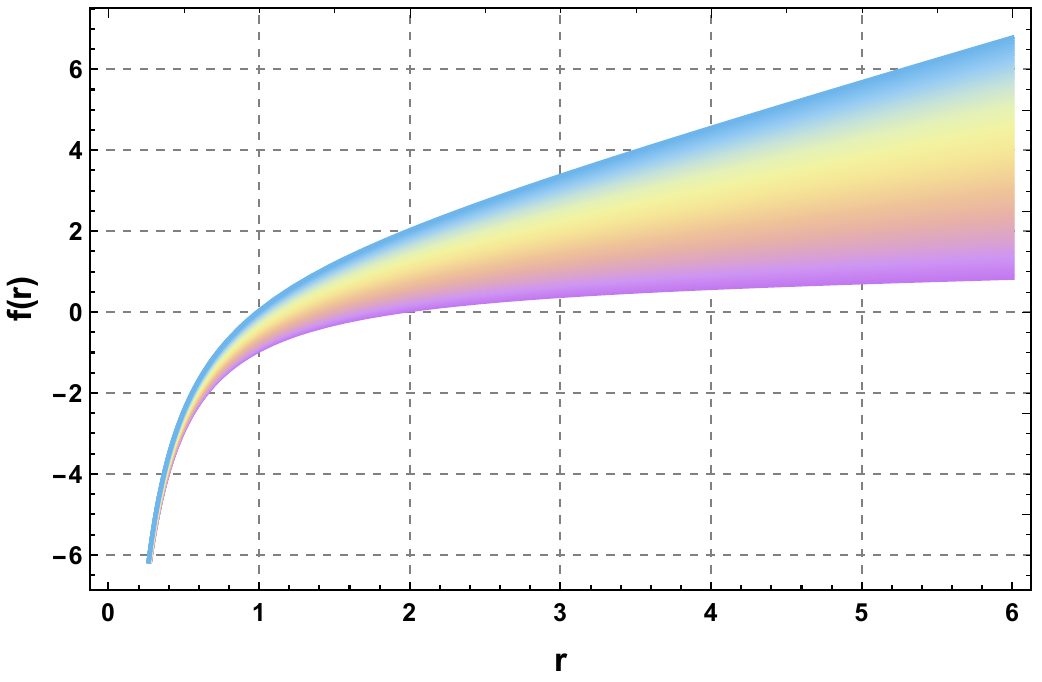}
    \includegraphics[height=5cm]{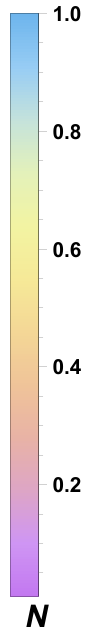}\qquad
    \includegraphics[height=5cm]{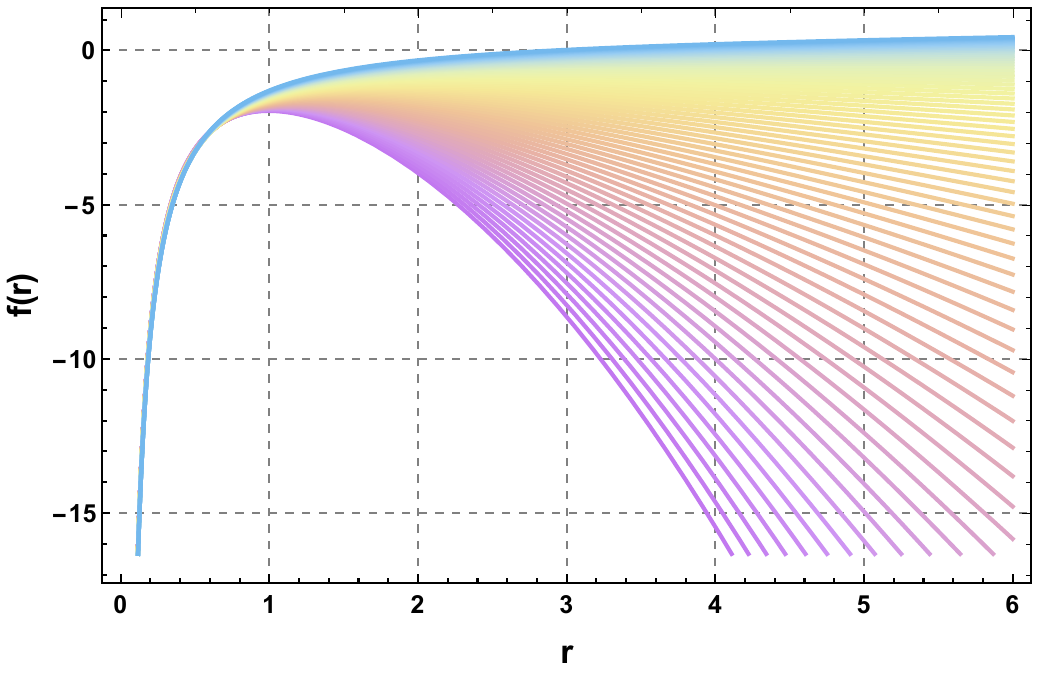}
    \includegraphics[height=5cm]{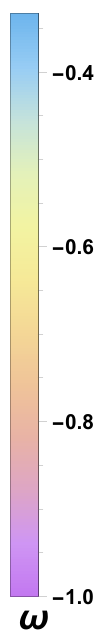} \\
    (c) $\alpha=\ell=0.1$, $\omega=-2/3$ and  $\Lambda=-0.01$ \hspace{3cm} (d) $\alpha=\ell=0.1$ and  $N=-\Lambda=0.01$ 
    \caption{ Plot of the metric function $f(r)$ with $M=1$.}
    \label{fig1}
\end{figure}

Figure~\ref{fig1} illustrates the behavior of the metric function $f(r)$ given in Eq.~\eqref{function} for $M=1$, highlighting the influence of the model parameters on the spacetime structure. In panel (a), varying the primary hair parameter $\ell$ produces small but noticeable deviations near the horizon, while preserving the asymptotic behavior. Panel (b) shows that the coupling constant $\alpha$ introduces an exponential correction that slightly shifts the profile of $f(r)$, as emphasized in the inset, indicating its relevance at short distances. In panel (c), the parameter $N$, associated with the surrounding field, significantly alters the growth of the metric function at larger radii, reflecting its contribution through the power-law term. Finally, panel (d) demonstrates the impact of the state parameter $\omega$, where different values lead to qualitatively distinct behaviors, including the possibility of rapid decrease of $f(r)$ for certain ranges, which may affect horizon formation. Overall, the combined effect of $(\alpha, \ell, \omega, N, \Lambda)$ leads to non-trivial modifications of the metric function, suggesting a rich horizon structure and potentially diverse thermodynamic properties.

\section{Thermodynamic Properties of the BH}\label{S4}

In this section, we will study the thermodynamic properties by
considering the static and spherically symmetric spacetime described by the line element \eqref{function}. As we discussed previously, this geometry incorporates several physically distinct contributions: the Schwarzschild term controlled by the mass parameter $M$, a surrounding field encoded by the state parameter $\omega$ and intensity $N$, an exponential correction governed by the coupling $\alpha$ and the primary hair scale $\ell$, and the cosmological constant $\Lambda$. The condition $f(r_h)=0$ determines the event horizon radius $r_h$ as the largest real root.

\subsection{BH Mass}\label{S4-1}

The gravitational mass can be expressed in terms of the horizon radius by imposing the horizon condition $f(r_+)=0$. To obtain the value of $M$, we need to solve the equation:
\begin{equation}\label{mass_new}
1-\frac{2M}{r_+}-\frac{N}{r_+^{3\omega+1}}+\alpha\,\exp\left(-\frac{r_+}{M-\frac{\alpha\ell}{2}}\right)-\frac{\Lambda}{3}r_+^{2}=0.
\end{equation}

Due to the non-linear dependence of the exponential term on $M$, this expression generally requires numerical treatment or perturbative expansion for explicit evaluation. This feature reflects the non-trivial backreaction induced by the primary hair parameter.

\begin{figure}[ht!]
    \centering
    \includegraphics[height=5cm]{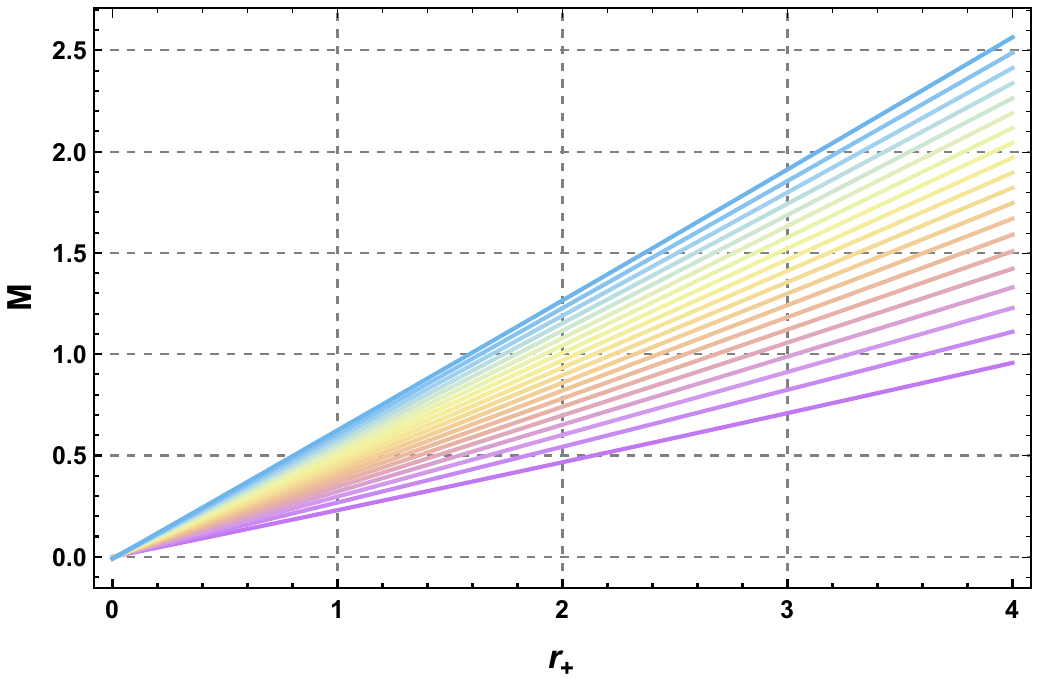}
    \includegraphics[height=5cm]{fig0a.pdf}\qquad
    \includegraphics[height=5cm]{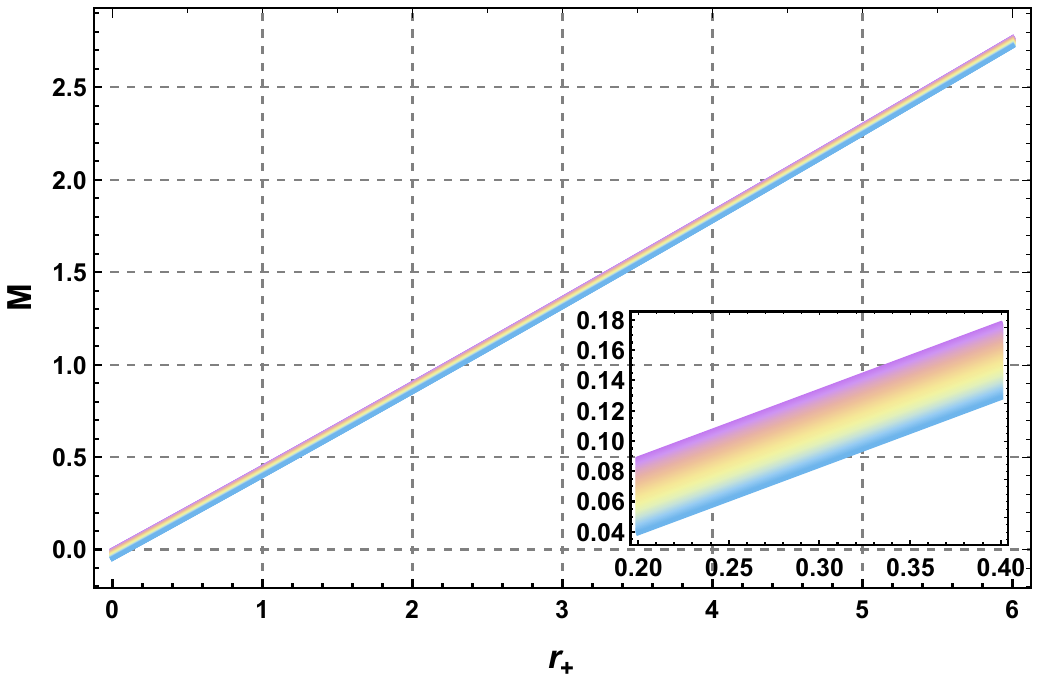}
    \includegraphics[height=5cm]{fig0b.pdf}\\
    (a) $\ell=0.1$, $\omega=-2/3$ and  $N=-\Lambda=0.01$ \hspace{3cm} (b) $\alpha=0.1$, $\omega=-2/3$ and  $N=-\Lambda=0.01$\\
    \includegraphics[height=5cm]{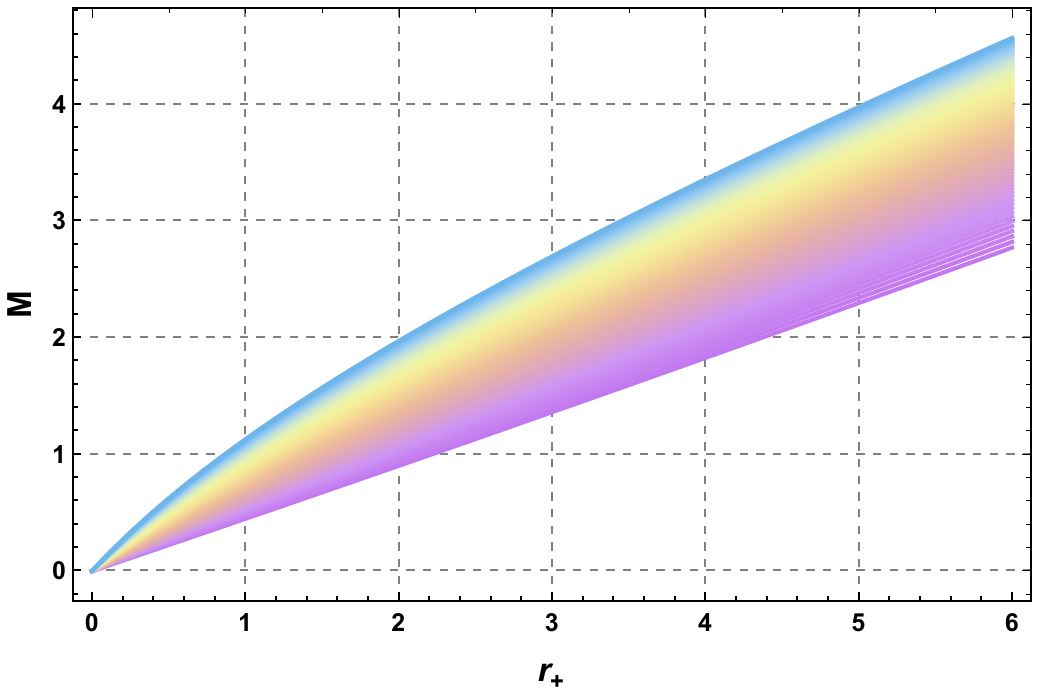}
    \includegraphics[height=5cm]{fig0c.pdf}\qquad
    \includegraphics[height=5cm]{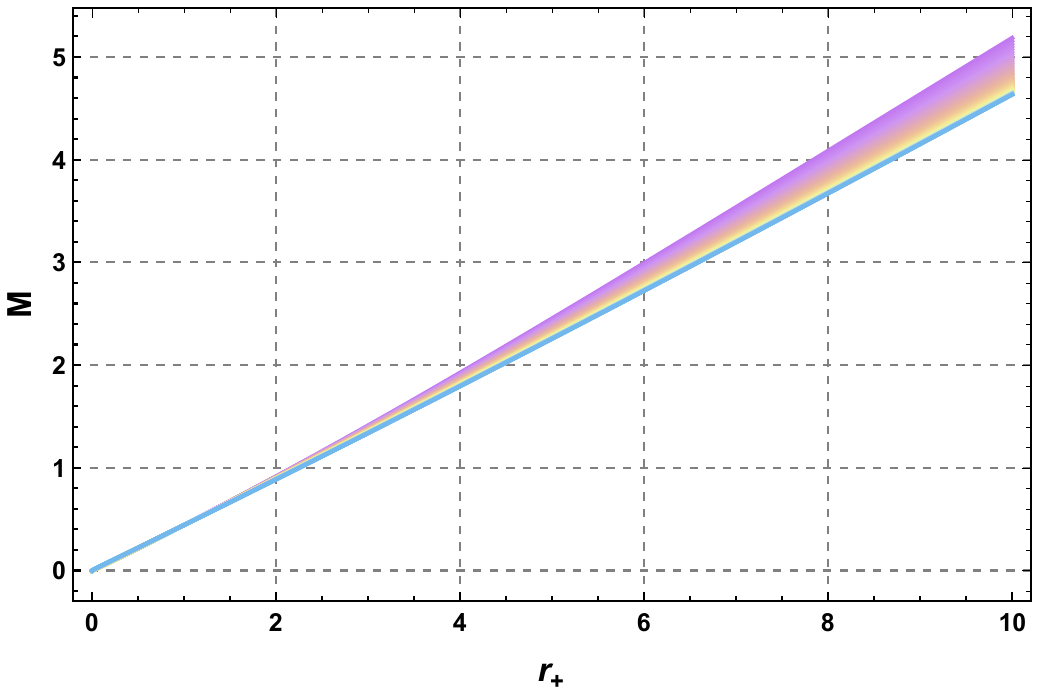}
    \includegraphics[height=5cm]{fig0d.pdf} \\
    (c) $\alpha=\ell=0.1$, $\omega=-2/3$ and  $\Lambda=-0.01$ \hspace{3cm} (d) $\alpha=\ell=0.1$ and  $N=-\Lambda=0.01$ 
    \caption{ Plot of the BH Mass.}
    \label{fig2}
\end{figure}

Figure~\ref{fig2} depicts the behavior of the black hole mass $M$, obtained from the horizon condition in Eq.~\eqref{mass_new}, as a function of the horizon radius $r_+$. In panel (a), variations in the primary hair parameter $\ell$ lead to noticeable changes in the slope of the $M$--$r_+$ relation, indicating that the hair contribution enhances the mass for larger values of $\ell$. Panel (b) shows that the coupling constant $\alpha$ introduces only mild corrections to the mass profile, as highlighted in the inset, suggesting that its effect remains subdominant for the chosen parameter range. In panel (c), the parameter $N$, associated with the surrounding field, significantly modifies the growth of the mass with $r_+$, producing a broader spread of curves and indicating a strong dependence on the field intensity. Finally, panel (d) illustrates the influence of the state parameter $\omega$, where different values lead to distinct linear behaviors, reflecting its role in controlling the effective matter contribution. Overall, the mass function exhibits a monotonic increase with the horizon radius, while the combined effects of $(\alpha, \ell, \omega, N, \Lambda)$ introduce non-trivial deviations that may impact the thermodynamic stability of the system.

\subsection{Hawking Temperature}\label{S4-2}

The Hawking temperature is determined from the surface gravity at the horizon, $\kappa=\tfrac{1}{2}f'(r_+)$, leading to
\begin{equation}\label{temp_new}
T=\frac{f'(r_+)}{4\pi}.
\end{equation}

By differentiating Eq.~\eqref{function}, we find
\begin{equation}
f'(r)=\frac{2M}{r^{2}}+\frac{(3\omega+1)N}{r^{3\omega+2}}+\alpha\,\frac{\exp\left(-\frac{r}{M-\frac{\alpha\ell}{2}}\right)}{M-\frac{\alpha\ell}{2}}-\frac{2\Lambda}{3}r.
\end{equation}

Thus, the Hawking temperature becomes
\begin{equation}\label{temp_explicit}
T=\frac{1}{4\pi}\left[\frac{2M}{r_+^{2}}+\frac{(3\omega+1)N}{r_+^{3\omega+2}}+\alpha\,\frac{\exp\left(-\frac{r_+}{M-\frac{\alpha\ell}{2}}\right)}{M-\frac{\alpha\ell}{2}}-\frac{2\Lambda}{3}r_+\right].
\end{equation}

This expression reveals that the exponential correction introduces a non-trivial modification to the temperature profile, which may significantly affect the evaporation process depending on the sign and magnitude of $\alpha$.

\begin{figure}[ht!]
    \centering
    \includegraphics[height=5cm]{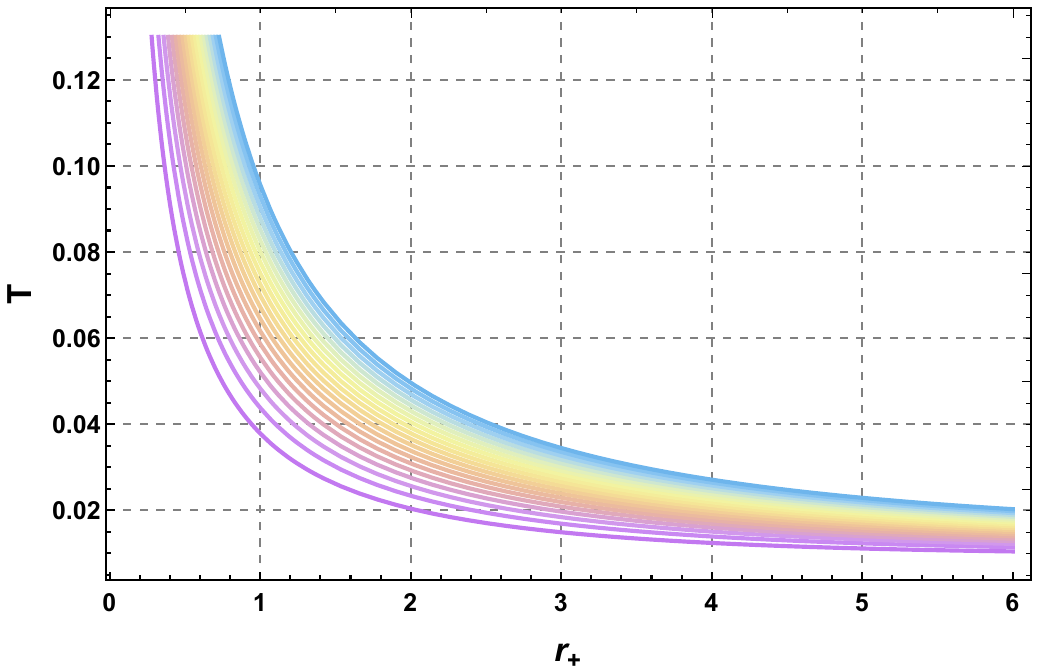}
    \includegraphics[height=5cm]{fig0a.pdf}\qquad
    \includegraphics[height=5cm]{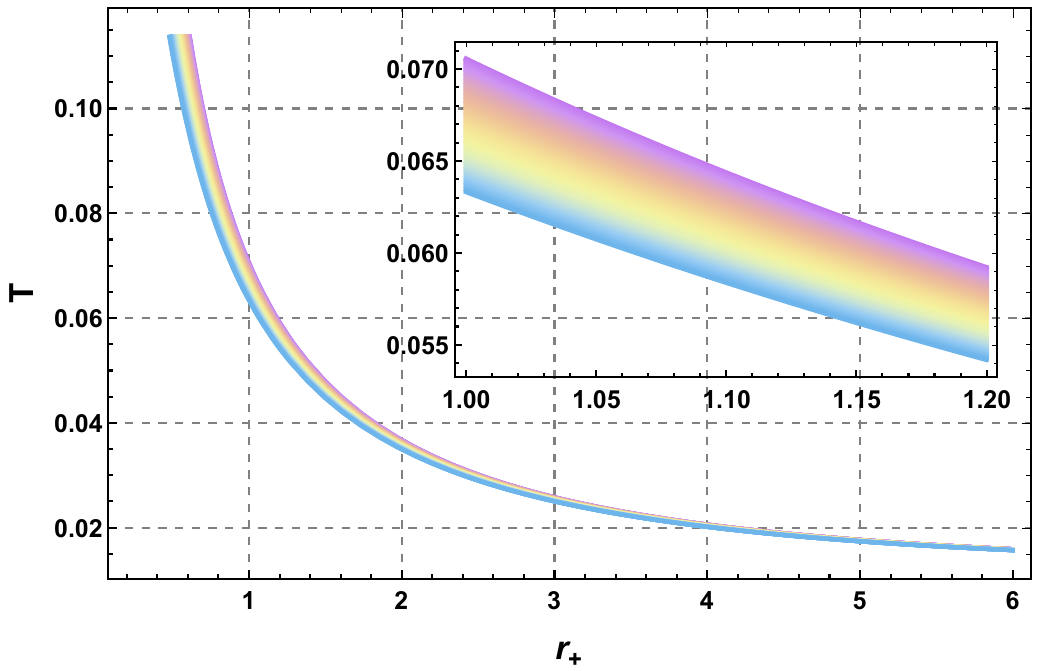}
    \includegraphics[height=5cm]{fig0b.pdf}\\
    (a) $\ell=0.1$, $\omega=-2/3$ and  $N=-\Lambda=0.01$ \hspace{3cm} (b) $\alpha=0.1$, $\omega=-2/3$ and  $N=-\Lambda=0.01$\\
    \includegraphics[height=5cm]{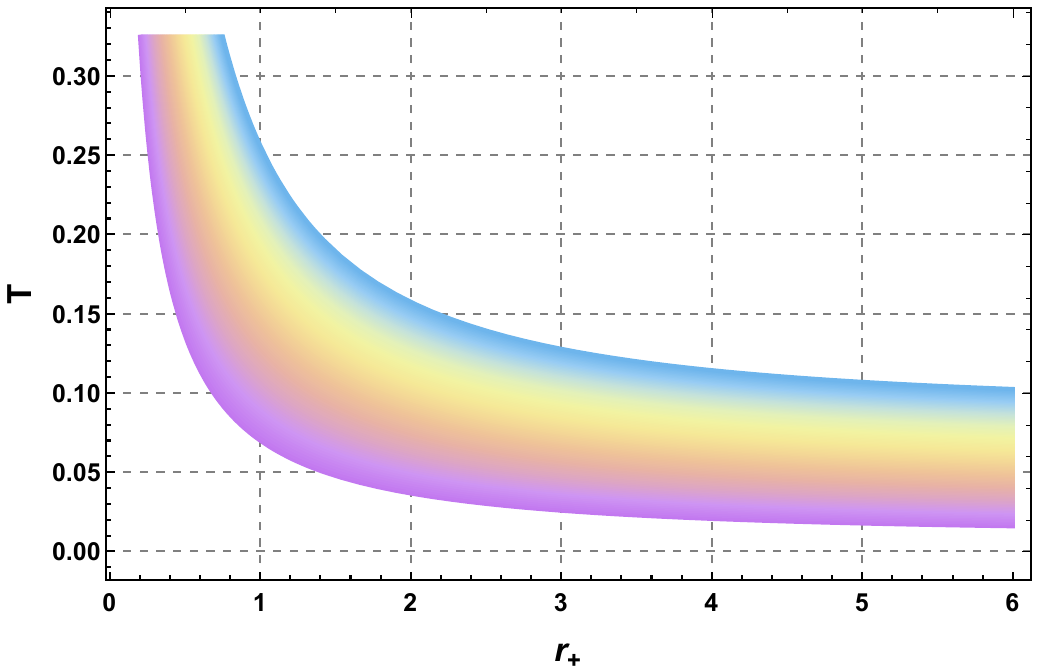}
    \includegraphics[height=5cm]{fig0c.pdf}\qquad
    \includegraphics[height=5cm]{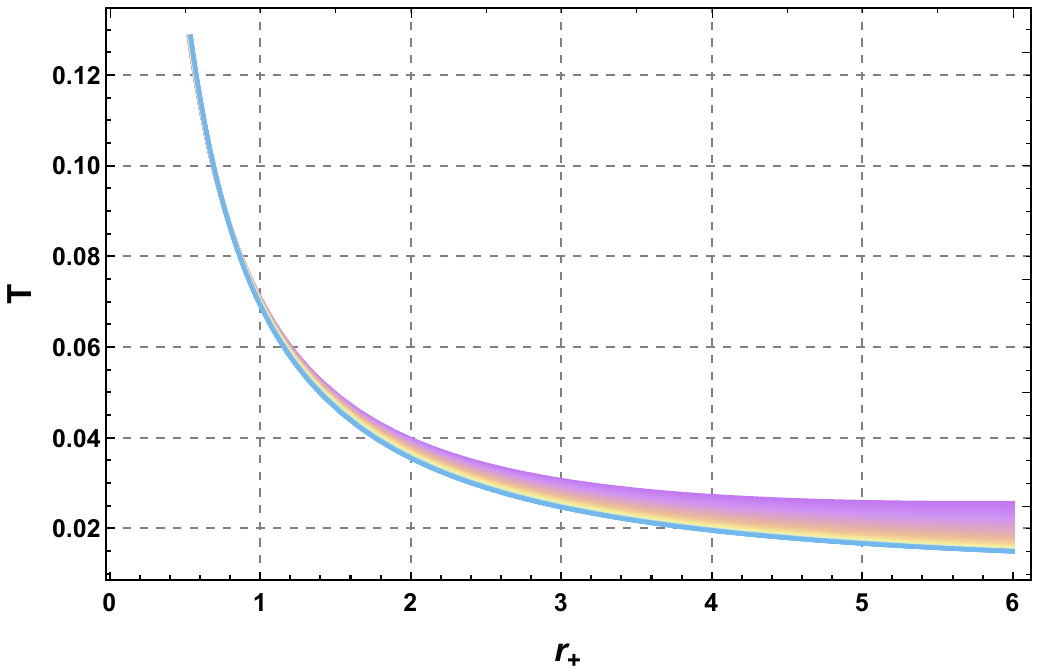}
    \includegraphics[height=5cm]{fig0d.pdf} \\
    (c) $\alpha=\ell=0.1$, $\omega=-2/3$ and  $\Lambda=-0.01$ \hspace{3cm} (d) $\alpha=\ell=0.1$ and  $N=-\Lambda=0.01$ 
    \caption{ Plot of the Hawking temperature.}
    \label{fig3}
\end{figure}

Figure~\ref{fig3} shows the behavior of the Hawking temperature $T$, given by Eq.~\eqref{temp_explicit}, as a function of the horizon radius $r_+$. In panel (a), the variation of the primary hair parameter $\ell$ modifies the temperature profile mainly at small radii, leading to a noticeable spread of curves and indicating its influence on the near-horizon geometry. Panel (b) illustrates that the coupling constant $\alpha$ introduces only slight temperature corrections, as highlighted in the inserted figure, confirming that its exponential contribution remains subdominant in the regime considered. In panel (c), the parameter $N$, associated with the surrounding field, significantly enhances the temperature for larger values, demonstrating its strong impact on the thermodynamic behavior. Finally, panel (d) displays the effect of the state parameter $\omega$, where different choices produce small but systematic deviations in the temperature profile. In all cases, the temperature decreases monotonically with increasing $r_+$, which is characteristic of asymptotically AdS black holes, while the combined effects of $(\alpha, \ell, \omega, N, \Lambda)$ introduce non-trivial corrections that may influence the evaporation process and thermal stability.

\subsection{Heat Capacity}\label{S4-3}

The local thermodynamic stability can be analyzed through the heat capacity
\begin{equation}
C = \frac{\partial M}{\partial T}
  = \frac{\partial M/\partial r_+}{\partial T/\partial r_+}.
\end{equation}

Given the implicit form of $M$ and the exponential dependence in Eq.~\eqref{mass_new}, the heat capacity acquires a highly non-linear structure. In particular, divergences of $C$ signal possible phase transitions, which are now influenced not only by the surrounding field $(\omega, N)$ and cosmological constant $\Lambda$, but also by the hair parameters $(\alpha, \ell)$. These additional contributions can shift critical points and modify stability regions in a non-trivial manner.

\begin{figure}[ht!]
    \centering
    \includegraphics[height=5cm]{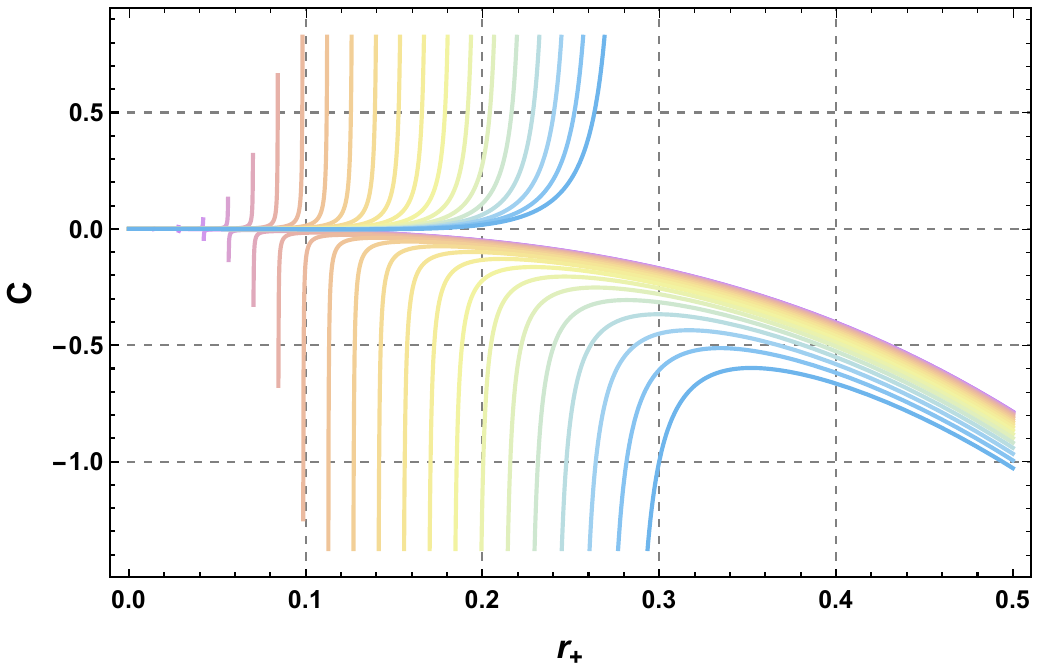}
    \includegraphics[height=5cm]{fig0a.pdf}\qquad
    \includegraphics[height=5cm]{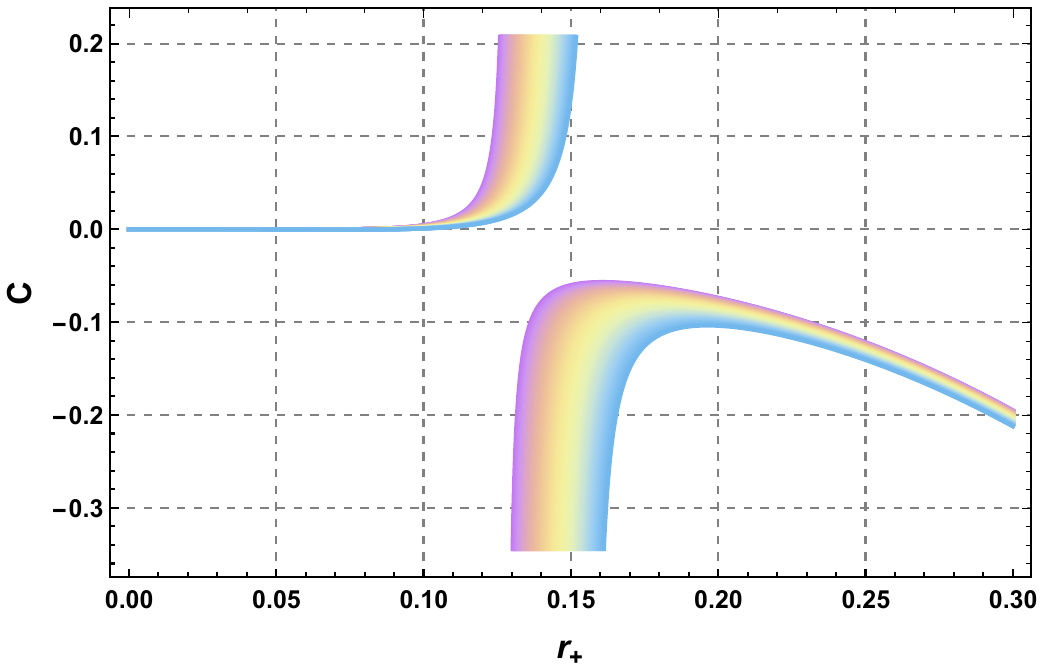}
    \includegraphics[height=5cm]{fig0b.pdf}\\
    (a) $\ell=0.1$, $\omega=-2/3$ and  $N=-\Lambda=0.01$ \hspace{3cm} (b) $\alpha=0.1$, $\omega=-2/3$ and  $N=-\Lambda=0.01$\\
    \includegraphics[height=5cm]{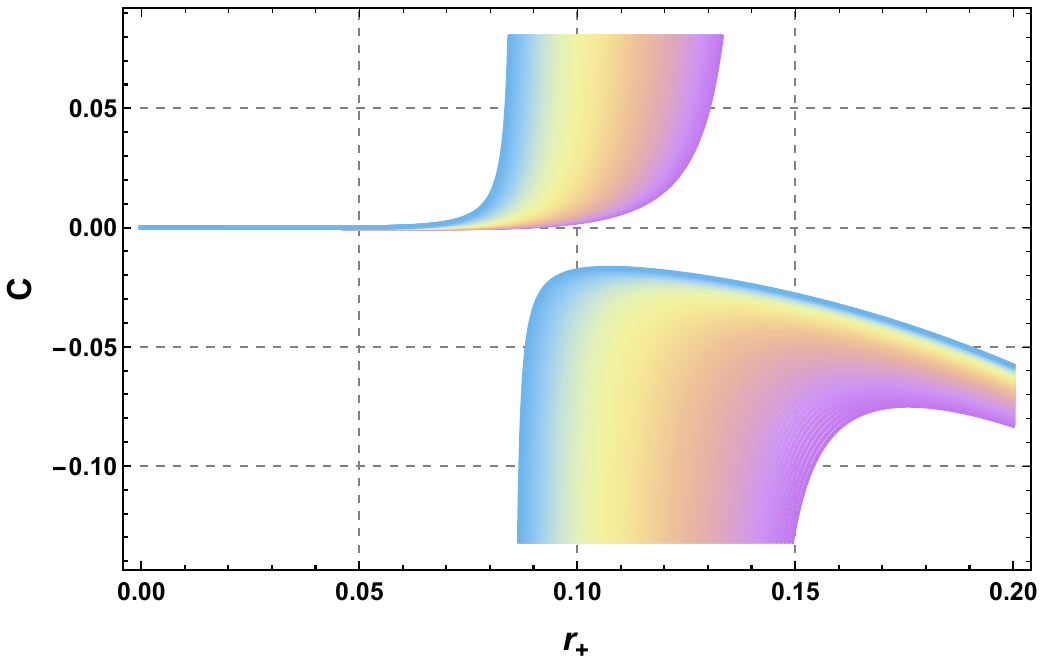}
    \includegraphics[height=5cm]{fig0c.pdf}\qquad
    \includegraphics[height=5cm]{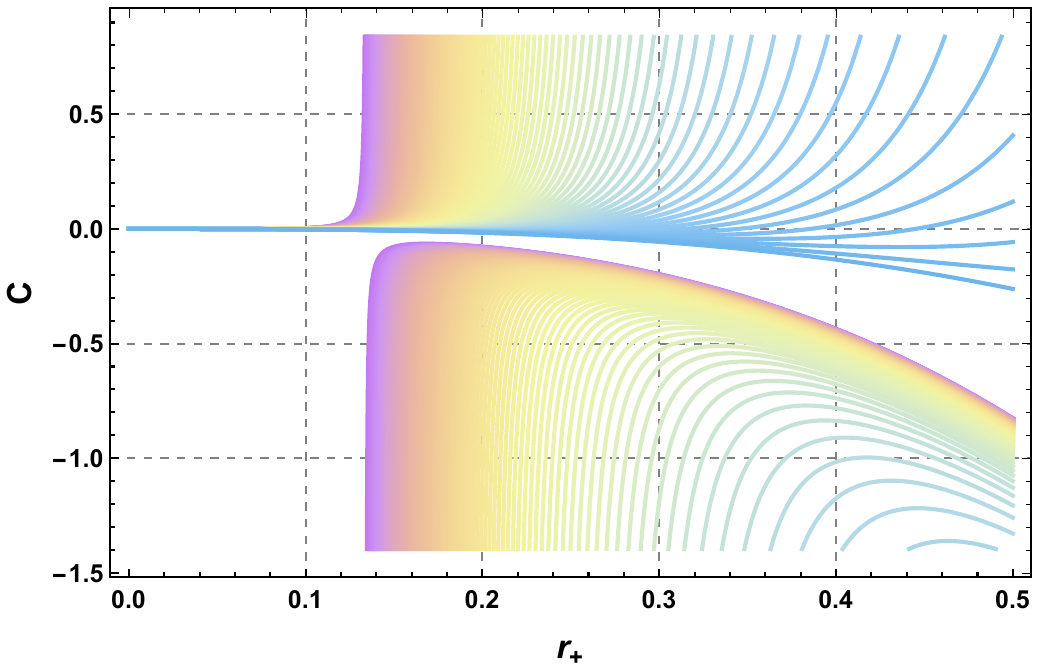}
    \includegraphics[height=5cm]{fig0d.pdf} \\
    (c) $\alpha=\ell=0.1$, $\omega=-2/3$ and  $\Lambda=-0.01$ \hspace{3cm} (d) $\alpha=\ell=0.1$ and  $N=-\Lambda=0.01$ 
    \caption{ Plot of the Heat Capacity.}
    \label{fig4}
\end{figure}

Figure~\ref{fig4} illustrates the behavior of the heat capacity $C$ as a function of the horizon radius $r_+$ for different choices of the parameters $(\alpha,\ell,\omega,N,\Lambda)$. In all panels, the heat capacity exhibits divergences at specific values of $r_+$, signaling second-order phase transitions between thermodynamically unstable ($C<0$) and stable ($C>0$) black hole configurations. Panel (a) shows that varying the primary hair parameter $\ell$ shifts the location of these critical points, modifying the size at which the transition occurs. In panel (b), the coupling constant $\alpha$ introduces significant deformations in the profile of $C$, enhancing the non-linear behavior induced by the exponential correction in the metric function. Panel (c) demonstrates that the combined effect of $\alpha$ and $\ell$ leads to a richer phase structure, including multiple branches of stability and instability. Finally, panel (d) highlights the role of the state parameter $\omega$, whose variation strongly affects both the magnitude and divergence structure of the heat capacity, indicating that the surrounding field plays a crucial role in determining the thermodynamic stability. Overall, the figure confirms that the interplay between the hair parameters and the external field significantly alters the phase transition pattern compared to standard black hole scenarios.

\subsection{Gibbs Free Energy}\label{S4-5}

For the present neutral configuration, the Gibbs free energy reduces to
\begin{equation}
G = M - T S,
\end{equation}
where the entropy follows the Bekenstein--Hawking area law, $S=\pi r_h^{2}$. Therefore,
\begin{equation}
G = M - \pi r_+^{2} T.
\end{equation}

The interplay between the exponential correction and the surrounding field modifies the global thermodynamic behavior, potentially leading to novel phase structures. In particular, the competition between the $\Lambda$-term and the hair-induced exponential contribution may generate multiple branches of solutions, indicating rich phase transition phenomena beyond the standard Schwarzschild--de Sitter scenario.

\begin{figure}[ht!]
    \centering
    \includegraphics[height=5cm]{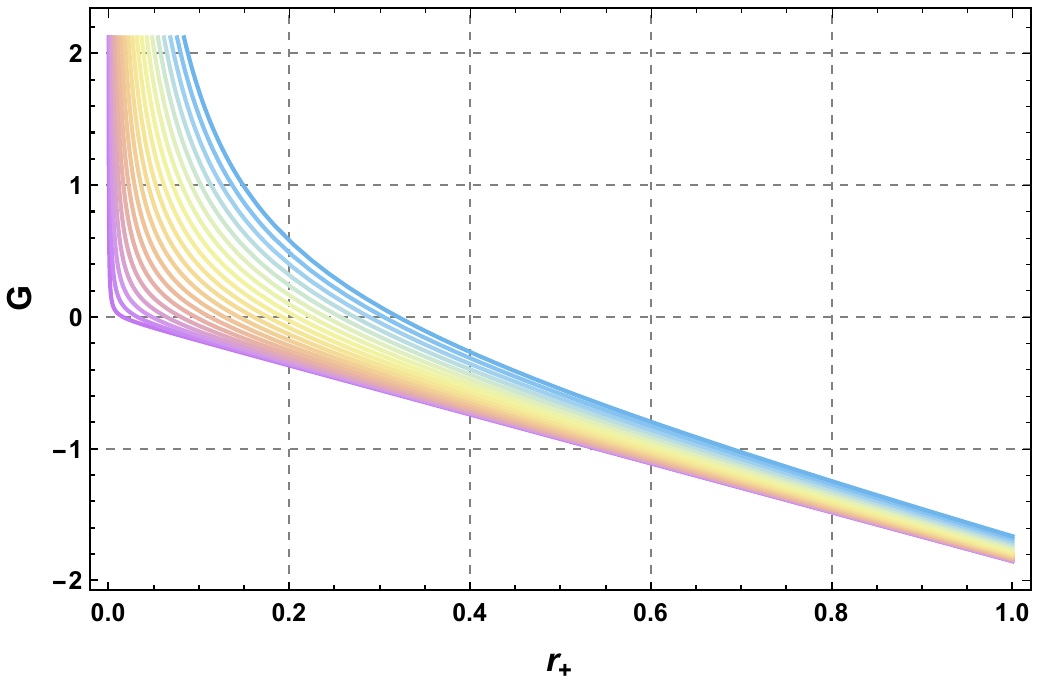}
    \includegraphics[height=5cm]{fig0a.pdf}\qquad
    \includegraphics[height=5cm]{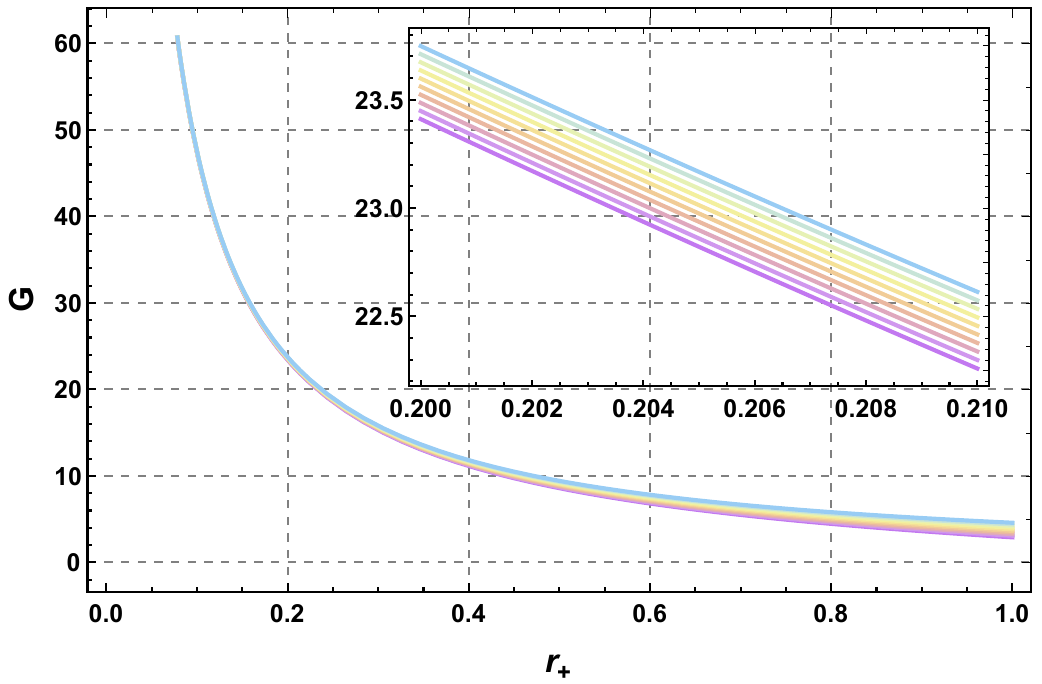}
    \includegraphics[height=5cm]{fig0b.pdf}\\
    (a) $\ell=0.1$, $\omega=-2/3$ and  $N=-\Lambda=0.01$ \hspace{3cm} (b) $\alpha=0.1$, $\omega=-2/3$ and  $N=-\Lambda=0.01$\\
    \includegraphics[height=5cm]{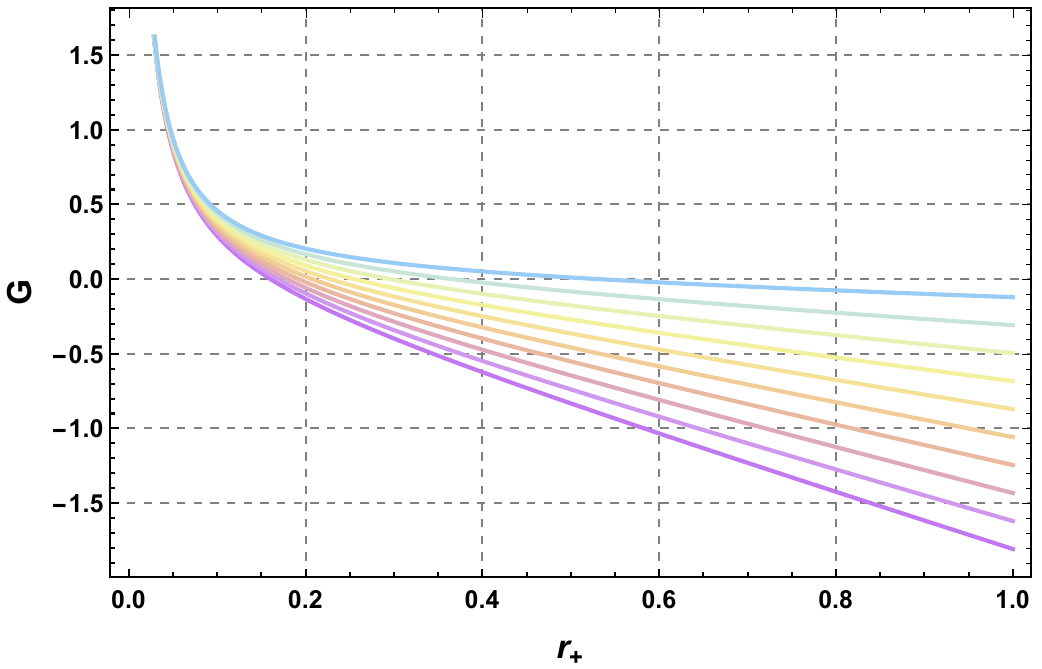}
    \includegraphics[height=5cm]{fig0c.pdf}\qquad
    \includegraphics[height=5cm]{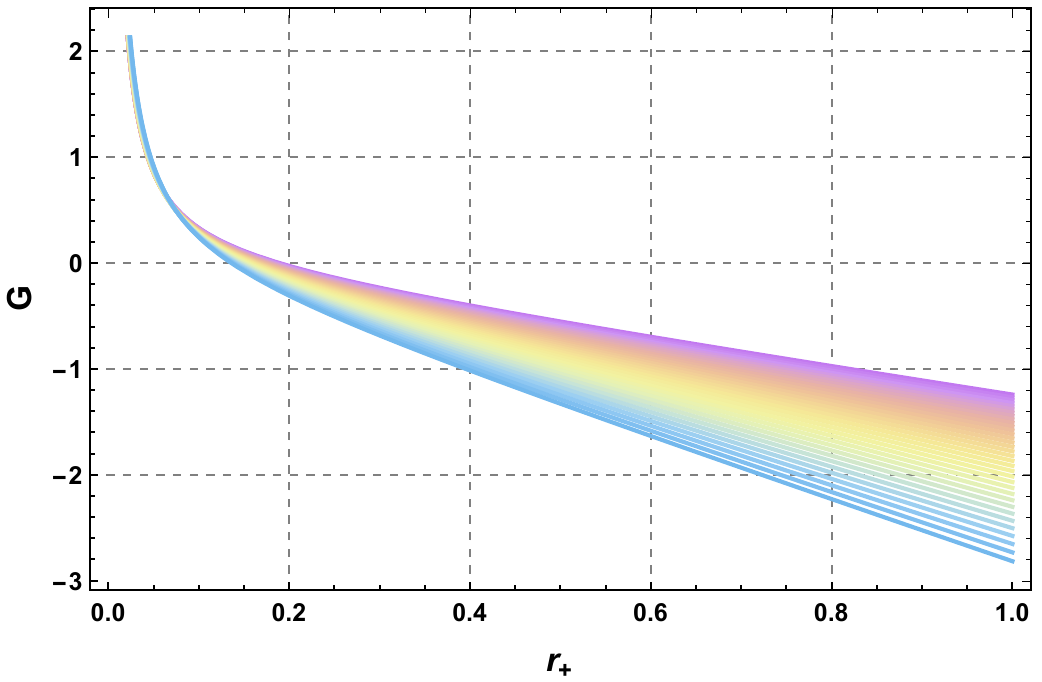}
    \includegraphics[height=5cm]{fig0d.pdf} \\
    (c) $\alpha=\ell=0.1$, $\omega=-2/3$ and  $\Lambda=-0.01$ \hspace{3cm} (d) $\alpha=\ell=0.1$ and  $N=-\Lambda=0.01$ 
    \caption{ Plot of the Gibbs Free Energy.}
    \label{fig5}
\end{figure}

Figure~\ref{fig5} displays the behavior of the Gibbs free energy $G$ as a function of the horizon radius $r_+$ for different choices of the parameters $(\alpha,\ell,\omega,N,\Lambda)$. In panel (a), the variation of the primary hair parameter $\ell$ shifts the free energy curves, indicating that the hair contribution modifies the global thermodynamic preference of the black hole states, particularly in the small-radius regime. Panel (b) shows that the coupling constant $\alpha$ affects the curvature of $G$, with the inset highlighting subtle differences in the intermediate region of $r_+$, suggesting the presence of competing thermodynamic phases. In panel (c), the influence of the surrounding field parameter $N$ becomes evident, as increasing its magnitude leads to a significant lowering of the Gibbs free energy, favoring more stable configurations. Finally, panel (d) illustrates the effect of the state parameter $\omega$, which alters both the slope and asymptotic behavior of $G$, thereby impacting the global stability and possible phase transitions. Overall, the figure indicates that the interplay between the hair parameters and the external field strongly modifies the thermodynamic landscape, potentially leading to rich phase structures beyond the standard AdS black hole scenario.

\section{Sparsity of Hawking Radiation}
\label{sec:sparsity}

In this section, we will analyze the sparsity of the Hawking flux emitted by the hairy Kiselev black hole surrounded by quintessence. The notion of sparsity provides a useful way to quantify how continuous, or instead how intermittent, the Hawking cascade is. Although Hawking radiation is often described through a thermal spectrum, the actual emission process is not a continuous classical flux. It consists of individual quanta emitted with a certain average time separation. The radiation is said to be sparse when the average time between successive quanta is much larger than the characteristic oscillation time, or localization time, of a typical emitted quantum.

For a bosonic field emitted by the black hole, the spectral energy flux can be written as follows
\begin{align}
\frac{d^2E}{dt\,d\varpi}
=\frac{1}{2\pi}
\sum_{l=0}^{\infty}
(2l+1)\,\Gamma_l(\varpi)\,
\frac{\varpi}{\exp(\varpi/T_H)-1},
\label{eq:energy_flux_sparsity}
\end{align}
where $\varpi$ denotes the frequency of the emitted quantum, introduced here to avoid confusion with the quintessence parameter $\omega_q$. The quantity $\Gamma_l(\varpi)$ is the greybody factor, which measures the probability that a mode produced near the horizon crosses the curvature potential barrier and reaches the asymptotic region. The corresponding number flux is obtained by dividing the energy spectrum by the energy $\varpi$ of each emitted quantum, namely
\begin{align}
\frac{d^2N}{dt\,d\varpi}=\frac{1}{2\pi}
\sum_{l=0}^{\infty}
(2l+1)\,\frac{\Gamma_l(\varpi)}{\exp(\varpi/T_H)-1}.
\label{eq:number_spectrum_sparsity}
\end{align}
Therefore, the total particle emission rate reads
\begin{align}
\dot{N}=\frac{1}{2\pi}
\sum_{l=0}^{\infty}
(2l+1)\int_0^\infty
\frac{\Gamma_l(\varpi)}{\exp(\varpi/T_H)-1}\,d\varpi.
\label{eq:Ndot_exact_sparsity}
\end{align}
The average time gap between successive emitted quanta is given by
\begin{align}
\tau_{\rm gap}=\frac{1}{\dot{N}} .
\label{eq:taugap_sparsity}
\end{align}
This quantity is the first ingredient in the sparsity parameter. A small value of $\dot{N}$ gives a large $\tau_{\rm gap}$, meaning that the black hole emits quanta only rarely. Conversely, a large number flux makes the Hawking cascade closer to an ordinary continuous thermal flux. The second ingredient is the characteristic timescale associated with a typical emitted quantum. A natural choice is the oscillation period of a quantum whose frequency is near the maximum of the spectrum, namely
\begin{align}
\tau_{\rm loc}=\frac{2\pi}{\varpi_\star},
\label{eq:tauloc_sparsity}
\end{align}
where $\varpi_\star$ may be chosen as the peak frequency of either the number spectrum or the energy spectrum. In an ideal blackbody approximation, these peaks are proportional to the Hawking temperature,
$\varpi_{\rm peak}^{(N)} = x_N T_H$ and $\varpi_{\rm peak}^{(E)} = x_E T_H$. For a three-dimensional Planck distribution, the dimensionless constants satisfy $2\left(1-e^{-x_N}\right)=x_N$ and $3\left(1-e^{-x_E}\right)=x_E$, giving $x_N \simeq 1.59362$ and $x_E \simeq 2.82144$ . The precise value of $\varpi_\star$ is modified once greybody factors are included, because $\Gamma_l(\varpi)$ suppresses the low-frequency and higher-angular-momentum parts of the spectrum. Nevertheless, the blackbody peak values provide a useful analytic estimate and allow one to see clearly how the black hole parameters affect sparsity.

In this context, the sparsity parameter is defined as the ratio between the average gap time and the localization time of the emitted quantum. Then, we define
\begin{align}
\eta=\frac{\tau_{\rm gap}}{\tau_{\rm loc}}=\frac{\varpi_\star}{2\pi \dot{N}}.
\label{eq:eta_exact_sparsity}
\end{align}
The interpretation of this expression is simple. If $\eta\ll1$, many quanta overlap within one characteristic oscillation time, and the radiation behaves approximately as a continuous flux. If $\eta\gg1$, the emission is highly intermittent: a quantum is emitted, then the black hole waits for a long time before emitting the next one. For four-dimensional black holes, the Hawking cascade is typically sparse, and this remains true even more strongly when greybody suppression is taken into account. Using Eq.~\eqref{eq:Ndot_exact_sparsity}, the exact greybody-corrected sparsity may be written as follows
\begin{align}
\eta=\frac{\varpi_\star}
{\displaystyle
\sum_{l=0}^{\infty}
(2l+1)\int_0^\infty
\frac{\Gamma_l(\varpi)}{\exp(\varpi/T_H)-1}\,d\varpi}.
\label{eq:eta_greybody_sparsity}
\end{align}
This expression shows that the sparsity is controlled by two competing effects. The temperature sets the typical energy scale of the emitted quanta, while the greybody factors determine how efficiently those quanta escape through the effective potential barrier. Since $0\leq \Gamma_l(\varpi)\leq 1$, greybody factors reduce the number flux
relative to a perfect blackbody and therefore increase $\eta$. Thus, the greybody-corrected Hawking radiation is always more sparse than the corresponding idealized Planckian estimate.

To obtain an analytic estimate, one may replace the full greybody spectrum by an effective blackbody emitter with area $\mathcal{A}_{\rm eff}$. In the geometric-optics limit this effective area is naturally related to the high-frequency absorption cross-section,
\begin{align}
\mathcal{A}_{\rm eff}\equiv \sigma_{\rm geo}=\pi b_c^2 ,
\label{eq:Aeff_geo}
\end{align}
where $b_c$ is the critical impact parameter of null geodesics. For the metric \eqref{metric}, the photon sphere radius $r_{\rm ph}$ is obtained from
\begin{align}
r_{\rm ph} f'(r_{\rm ph})-2f(r_{\rm ph})=0 ,
\label{eq:photon_sphere_sparsity}
\end{align}
and the corresponding critical impact parameter reads
\begin{align}
b_c^2=\frac{r_{\rm ph}^2}{f(r_{\rm ph})}.
\label{eq:bc_sparsity}
\end{align}
Therefore, we have
\begin{align}
\mathcal{A}_{\rm eff}=\pi \frac{r_{\rm ph}^2}{f(r_{\rm ph})}.
\label{eq:Aeff_sparsity}
\end{align}
For comparison, one may also use the horizon area $A_h=4\pi r_h^2$.
The geometric-optics area is usually more appropriate for high-frequency quanta, while the horizon area gives a simpler estimate of the emission scale. In the blackbody approximation, the total number flux takes the form
\begin{align}
\dot{N}_{\rm bb}=\frac{g\,\zeta(3)}{4\pi^2}\,
\mathcal{A}_{\rm eff} T_H^3 ,
\label{eq:Ndot_bb_sparsity}
\end{align}
where $g$ counts the number of radiating degrees of freedom and $\zeta(3)$ is the Riemann zeta function. Substituting this expression into Eq.~\eqref{eq:eta_exact_sparsity} yields
\begin{align}
\eta_{\rm bb}^{(\star)}=\frac{2\pi x_\star}
{g\,\zeta(3)\,\mathcal{A}_{\rm eff}T_H^2},
\label{eq:eta_bb_sparsity}
\end{align}
with $x_\star= \frac{\varpi_\star}{T_H}$. Equivalently, using the thermal wavelength, we obtain
\begin{align}
\lambda_{\rm th}=\frac{2\pi}{T_H},
\end{align}
one can write
\begin{align}
\eta_{\rm bb}^{(\star)}=\frac{x_\star}{2\pi g\zeta(3)}
\frac{\lambda_{\rm th}^2}{\mathcal{A}_{\rm eff}} .
\label{eq:eta_lambda_sparsity}
\end{align}
This expression makes the physical meaning transparent: the Hawking flux is sparse whenever the square of the typical wavelength of the emitted quantum is much larger than the effective emitting area. Since black holes radiate with wavelengths of order $1/T_H$, and since $T_H$ is fixed by the surface gravity, sparsity is ultimately a near-horizon
geometric effect. Using Eq.~\eqref{eq:eta_lambda_sparsity}, the blackbody sparsity can be expressed directly in
terms of the metric function as follows
\begin{align}
\eta_{\rm bb}^{(\star)}
=\frac{32\pi^3 x_\star}
{g\,\zeta(3)\,
\mathcal{A}_{\rm eff}
\left[f'(r_h)\right]^2}.
\label{eq:eta_metric_sparsity}
\end{align}
Therefore, for the hairy Kiselev geometry, we arrive at
\begin{align}
\eta_{\rm bb}^{(\star)}=
\frac{32\pi^3 x_\star}
{g\,\zeta(3)\,
\mathcal{A}_{\rm eff}
\left[\frac{2M}{r_h^2}
+(3\omega_q+1)N r_h^{-(3\omega_q+2)}
-\frac{\alpha}{\sigma}\exp\left(-\frac{r_h}{\sigma}\right)
-\frac{2\Lambda}{3}r_h
\right]^2}.
\label{eq:eta_explicit_sparsity}
\end{align}
This is the central analytic result of the sparsity analysis. It shows explicitly how the hair parameters, the surrounding quintessence, and the cosmological constant enter the intermittency of Hawking emission.

\subsection{Role of the hairy and quintessential parameters}

The sparsity parameter is inversely proportional to
$\mathcal{A}_{\rm eff}T_H^2$. Hence, any parameter that increases the temperature or enlarges the effective absorption area tends to make the radiation less sparse. Conversely, any parameter that cools the black hole or increases the potential barrier tends to make
the emission more sparse. The exponential hair contribution enters through the factor
\begin{align}
-\frac{\alpha}{\sigma}\exp\left(-\frac{r_h}{\sigma}\right),
\qquad
\sigma=M-\frac{\alpha\ell}{2}.
\end{align}
For $\sigma>0$ and positive $\alpha$, this term decreases $f'(r_h)$ at fixed horizon radius, thereby lowering the Hawking temperature and increasing $\eta$. Physically, the hair acts as a short-range deformation of the near-horizon geometry. Its influence is strongest for small black holes, where the horizon radius is comparable to the length scale $\sigma$. For large black holes, the exponential factor is suppressed, and the hair correction becomes less relevant. The primary hair parameter $\ell$ modifies the same contribution through
$\sigma$, so that changing $\ell$ shifts the strength and radial support of the exponential deformation. The quintessence contribution is governed by $(3\omega_q+1)N r_h^{-(3\omega_q+2)}$. For the usual quintessence interval $-1<\omega_q<-1/3$, the coefficient
$3\omega_q+1$ is negative. Therefore, depending on the sign and magnitude of $N$, this term may reduce the surface gravity at fixed $M$ and $r_h$. In that case the black hole becomes colder and the Hawking emission becomes more sparse. However, when the mass
is obtained implicitly from the horizon equation, the variation of $N$ also shifts $r_h$ and $\mathcal{A}_{\rm eff}$, so the net behavior must be read from the full expression \eqref{eq:eta_explicit_sparsity}. In general, the surrounding fluid modifies both the thermal
scale and the scattering barrier, and therefore it affects sparsity through the temperature
and through the greybody factors.

The cosmological constant contributes as follows
\begin{align}
-\frac{2\Lambda}{3}r_h .
\end{align}
For anti-de Sitter configurations, $\Lambda<0$, this term is positive and becomes more important for large black holes. It tends to increase the Hawking temperature, thereby reducing the sparsity. This is consistent with the known behavior of large AdS black holes,
which can radiate more efficiently. For de Sitter configurations, $\Lambda>0$, the same term decreases the event-horizon surface gravity. Moreover, the physical region is bounded by a cosmological horizon, and the emission process must be treated between the two horizons. In that case, the finite size of the static patch and the cosmological redshift can strongly affect the emitted spectrum.

\subsection{Greybody correction and sparsity enhancement}

The blackbody estimate gives the minimal qualitative structure, but the true emission is filtered by the effective potential felt by perturbations. For a massless scalar field, the radial perturbation equation can be written in Schrödinger-like form, namely
\begin{align}
\frac{d^2u}{dr_\ast^2}+\left[
\varpi^2-V_{\rm eff}(r)\right]u=0,
\end{align}
where $dr_\ast=dr/f(r)$ and $V_{\rm eff}(r)$ is the effective potential written as follows
\begin{align}
V_{\rm eff}(r)=f(r)\left[\frac{l(l+1)}{r^2}+\frac{f'(r)}{r}\right].
\label{eq:Veff_sparsity}
\end{align}
In this context, the greybody factor $\Gamma_l(\varpi)$ is the transmission probability through this potential. Since the potential barrier becomes higher for larger $l$, higher multipoles are suppressed more strongly. Thus, the dominant contribution to the Hawking number flux usually comes from the lowest modes, especially the $l=0$ scalar mode. In the next section, we will examine the greybody factor for scalar perturbation with more details. Here, let us discuss the sparsity. Here, it is easy to note that the presence of the hair and quintessence terms changes the height and width of $V_{\rm eff}$. A higher and wider barrier lowers $\Gamma_l(\varpi)$, reduces $\dot{N}$, and consequently increases $\eta$. A lower barrier has the opposite effect. Hence, the greybody-corrected sparsity satisfies schematically
\begin{align}
\eta_{\rm gb}=\frac{\varpi_\star}{2\pi \dot{N}_{\rm gb}}
\geq\eta_{\rm bb},
\end{align}
whenever the blackbody estimate is taken with the same effective emitting area and $\Gamma_l(\varpi)\leq 1$. This inequality expresses the simple physical fact that scattering outside the horizon makes the already intermittent Hawking cascade even more sparse.

To conclude this section, let us discuss the resulting physical picture, which is clear. In this conjecture, we demonstrated that the sparsity of the Hawking radiation from the hairy Kiselev black hole is mainly controlled by the combination $\mathcal{A}_{\rm eff}T_H^2$.
The temperature determines the typical frequency and wavelength of the emitted quanta, whereas the effective area determines how many modes can be emitted efficiently. Since the hairy and quintessential parameters modify both the horizon structure and the external potential barrier, they leave direct imprints on the intermittency of the Hawking flux. For small black holes, the temperature is relatively high and the typical wavelength is shorter. This reduces the sparsity, although the flux may still remain far from continuous. For larger black holes, the temperature generally decreases, the thermal wavelength grows, and the average time interval between successive emissions increases. The radiation then becomes increasingly sparse. The exponential hair correction is most relevant near the small-radius regime, while the quintessence and cosmological terms become especially
important at larger distances. Therefore, the model naturally allows different sparsity profiles depending on whether the near-horizon hair or the large-scale cosmological fluid dominates the geometry.

With the analysis performed in this section, we note that the hairy Kiselev black hole emits Hawking radiation in a highly intermittent manner. The parameters $\alpha$ and $\ell$ encode short-range hairy corrections that can cool the black hole and increase sparsity, while $N$ and $\omega_q$ describe the surrounding quintessential matter and can either enhance or suppress emission depending on their effect on the surface gravity and greybody barrier. The cosmological constant further separates the behavior of AdS and dS configurations. The compact formula \eqref{eq:eta_explicit_sparsity} provides a direct way to evaluate these effects once the horizon radius and the effective absorption area are determined.

\section{Massless Scalar Perturbations around a General Static Spherically Symmetric Black Hole}

In this section, we provide a complete analysis of massless scalar field perturbations propagating in the background of the static spherically symmetric black hole described by the line element \eqref{metric}. In this context, massless scalar perturbations are of fundamental importance: they serve as test fields for classical stability, encode the ringdown phase of gravitational waves via quasinormal modes (QNMs), determine greybody factors and absorption cross-sections, and probe late-time tails. Since the background is static, spherically symmetric, and asymptotically (A)dS or flat depending on $\Lambda$, the analysis follows the standard Regge-Wheeler-Zerilli framework adapted to spin-$s=0$ fields.

The dynamic of a massless minimally coupled scalar field $\Phi$ is governed by the Klein-Gordon equation
\begin{equation}
\Box\Phi \equiv \frac{1}{\sqrt{-g}}\partial_\mu\left(\sqrt{-g}\,g^{\mu\nu}\partial_\nu\Phi\right) = 0.
\end{equation}
For the metric \eqref{metric}, $\sqrt{-g} = r^2\sin\theta$. Substituting the inverse metric components $g^{tt} = -1/f$, $g^{rr} = f$, $g^{\theta\theta} = 1/r^2$, and $g^{\phi\phi} = 1/(r^2\sin^2\theta)$, the equation becomes
\begin{equation}
\frac{1}{f}\frac{\partial^2\Phi}{\partial t^2} - \frac{1}{r^2}\partial_r\left(f r^2\partial_r\Phi\right) + \frac{1}{r^2}\Delta_{S^2}\Phi = 0,
\label{eq:KG}
\end{equation}
where $\Delta_{S^2}$ is the spherical Laplacian. We can now separate variables using the ansatz
\begin{equation}
\Phi(t,r,\theta,\phi) = e^{-i\omega t}\,Y_{lm}(\theta,\phi)\,\frac{u(r)}{r},
\end{equation}
where $Y_{lm}$ are spherical harmonics satisfying $\Delta_{S^2}Y_{lm} = -l(l+1)Y_{lm}$ ($l=0,1,2,\dots$ and $m=-l,\dots,l$), and $\omega$ is the (possibly complex) frequency. The factor $1/r$ is chosen for convenience in obtaining a Schrödinger-like form. Substituting into \eqref{eq:KG} and multiplying through by $r$ yields the radial master equation
\begin{equation}
f\frac{d^2u}{dr^2} + f'\frac{du}{dr} + \left(\frac{\omega^2}{f} - \frac{l(l+1)}{r^2} - \frac{f'}{r}\right)u = 0,
\label{eq:radial}
\end{equation}
where a prime denotes $d/dr$. To cast \eqref{eq:radial} into a one-dimensional wave equation, we introduce the tortoise coordinate
\begin{equation}
r_* = \int^r\frac{dr'}{f(r')},
\end{equation}
which satisfies $dr_*/dr = 1/f(r)$ (so $dr/dr_* = f(r)$ and $d/dr = f\,d/dr_*$). Near the event horizon $r_h$ (outermost positive root of $f(r_h)=0$), $r_*\to-\infty$; at spatial infinity, $r_*\to+\infty$ (or a finite value in pure de Sitter).

Changing variables in \eqref{eq:radial} (after lengthy but standard algebra involving the chain rule twice) eliminates the first-derivative term and produces the Schrödinger-like wave equation
\begin{equation}
\frac{d^2u}{dr_*^2} + \bigl(\omega^2 - V(r)\bigr)u(r_*) = 0,
\label{eq:schrodinger}
\end{equation}
where the effective potential is
\begin{equation}
V(r) = f(r)\left[\frac{l(l+1)}{r^2} + \frac{f'(r)}{r}\right].
\label{eq:V}
\end{equation}
This form is universal for massless scalar ($s=0$) perturbations in any static spherically symmetric metric of the form \eqref{metric}. For comparison, in Schwarzschild ($f=1-2M/r$), it reduces to the well-known $V(r) = f(r)[l(l+1)/r^2 + 2M/r^3]$.

Differentiating the given $f(r)$ (with $\sigma = M - \alpha\ell/2$),
\begin{align}
f'(r) &= \frac{2M}{r^2} + N(3\omega+1)r^{-(3\omega+2)} - \frac{\alpha}{\sigma}\exp\left(-\frac{r}{\sigma}\right) - \frac{2\Lambda r}{3}.
\end{align}
Thus, one has
\begin{align}
\frac{f'(r)}{r} &= \frac{2M}{r^3} + N(3\omega+1)r^{-(3\omega+3)} - \frac{\alpha}{\sigma r}\exp\left(-\frac{r}{\sigma}\right) - \frac{2\Lambda}{3}.
\end{align}
The effective potential therefore reads
\begin{equation}
V(r) = f(r)\left[\frac{l(l+1)}{r^2} + \frac{2M}{r^3} + N(3\omega+1)r^{-(3\omega+3)} - \frac{\alpha}{\sigma r}\exp\left(-\frac{r}{\sigma}\right) - \frac{2\Lambda}{3}\right].
\label{eq:Vexplicit}
\end{equation}
All parameters ($M$, $N$, $\omega$, $\alpha$, $\ell$, $\Lambda$, $l$) enter explicitly, allowing parametric studies of stability and spectra. Once we have obtained the effective potential, let us examine its asymptotic Behavior and potential shape. In this, we have the following situations
\begin{itemize}
\item \textbf{Near the horizon} ($r\to r_h^+$): $f(r)\sim f'(r_h)(r-r_h)$, so $V(r)\to 0$. The solutions behave as $u\sim e^{\pm i\omega r_*}$ (ingoing/outgoing waves).
\item \textbf{At spatial infinity}:
  \begin{itemize}
  \item If $\Lambda=0$ (asymptotically flat), $f(r)\to1$, $V(r)\sim l(l+1)/r^2 + \mathcal{O}(1/r^3)$ (power-law decay).
  \item If $\Lambda>0$ (de Sitter), $f(r)\sim -\Lambda r^2/3$, and the effective potential approaches a constant negative value; the spacetime is bounded and QNMs are discrete.
  \item If $\Lambda<0$ (anti-de Sitter), $f(r)\sim -\Lambda r^2/3$ with positive sign, leading to a confining potential; boundary conditions at infinity are Dirichlet-like.
  \end{itemize}
\item \textbf{Barrier structure}: For typical astrophysical parameters ($l\geq0$, small $N$, $\alpha$), $V(r)$ exhibits a positive peak outside $r_h$ (scattering barrier) and decays to zero at both ends. The height and width of the barrier control transmission/reflection coefficients. Negative regions (if any, e.g., large $\Lambda>0$ or exotic $\omega$) could signal instabilities, but for the parameter ranges where $f(r)>0$ outside $r_h$ and $V(r)\geq0$, the spacetime is classically stable against massless scalar perturbations.
\end{itemize}

Since $V(r)$ is real and the wave equation \eqref{eq:schrodinger} has the form of a time-independent Schrödinger equation with no negative-energy bound states (for physically motivated parameters where the potential barrier is positive), there are no exponentially growing modes. Time-domain evolution (via numerical integration of the wave equation) would show decaying oscillations, confirming stability. The presence of the quintessence term ($N$) and exponential correction ($\alpha$) generally raises or lowers the barrier height depending on signs, but does not introduce instabilities for the massless case.

The transmission coefficient (greybody factor) $\mathcal{T}(\omega,l) = |A_{\rm out}/A_{\rm in}|^2$ is obtained by solving the scattering problem on the potential $V(r)$. The absorption cross-section for low-frequency scalars approaches the horizon area. At late times ($t\to\infty$), power-law tails $\Phi\sim t^{-(2l+2)}$ appear due to backscattering off the $1/r^3$ tail of $V(r)$ (modified by the exponential and $\Lambda$ terms). It is important to highlight that we can calculate  rigorous lower bound on the greybody factor $|T_b|$ through the following expression
\begin{equation}
|T_b| \geq \mathrm{sech}^{2}\!\left( \int_{-\infty}^{+\infty} G \, dr_{*} \right),
\label{eq:greybody_bound}
\end{equation}
where we have defined the function $G$ as
\begin{equation}
G = \frac{\sqrt{\left(\xi'\right)^{2} + \left(\omega^{2} - V - \xi^{2}\right)^{2}}}{2\,\xi}.
\label{eq:G_def}
\end{equation}
Above, $\omega$ stands for the mode frequency, $V$ is the effective potential, $r_{*}$ is the tortoise coordinate, and $\xi(r_{*})$ is a strictly positive auxiliary function satisfying the asymptotic conditions
\begin{equation}
\xi(-\infty) = \xi(+\infty) = \omega.
\end{equation}
When $\xi$ is chosen to be constant and equal to $\omega$, the bound in Eq.~\eqref{eq:greybody_bound} simplifies considerably, yielding
\begin{equation}
|T_b| \geq \mathrm{sech}^{2}\!\left( \int_{-\infty}^{+\infty} \frac{V}{2\omega} \, dr_{*} \right)
\geq \mathrm{sech}^{2}\!\left( \int_{r_h}^{+\infty} \frac{V}{2\omega\, f(r)} \, dr \right),
\label{eq:greybody_simplified}
\end{equation}
where $r_h$ denotes the event horizon radius and $f(r)$ is the metric function relating the radial coordinate $r$ to the tortoise coordinate via $dr_{*} = dr/f(r)$.

The integral therefore becomes
\[
I = \int_{r_h}^{\infty} \left[ \frac{l(l+1)}{2\omega r^2} + \frac{f'(r)}{2\omega r} \right] dr,
\]
and the lower bound is \( |T_b| \geq \mathrm{sech}^2(I) \). This integral is finite (or evaluated to a large cutoff in non-asymptotically-flat cases) and can be evaluated numerically after determining the event horizon \(r_h\) (outermost root of \(f(r_h)=0\)). We fix \(M=1\) (frequencies scale as \(1/M\)) and \(\omega = 0.5\). We consider the four benchmark cases used in the QNM analysis:
\begin{itemize}
\item Schwarzschild: \(N=\alpha=\Lambda=0\)
\item Quintessence: \(N=0.05\), \(\omega=-2/3\)
\item Exponential correction: \(\alpha=0.05\), \(\ell=1\)
\item Cosmological constant: \(\Lambda=0.01\)
\end{itemize}

The resulting lower bounds on \(|T_b|\) are shown below. Note that for \(\Lambda>0\) (de Sitter) the integrand becomes negative at large \(r\), and the tortoise coordinate is finite up to the cosmological horizon; the values are obtained with a large but finite cutoff consistent with the physical region.

\begin{table}[ht]
\centering
\caption{Lower bounds on the greybody factor \(|T_b|\) for \(\omega=0.5\), \(M=1\).}
\begin{tabular}{c|cc|cc}
Case & \(l=0\) & Bound & \(l=1\) & Bound \\
\hline
Schwarzschild & \(I=0.250\) & \(0.940\) & \(I=1.250\) & \(0.281\) \\
Quintessence   & \(I=-0.223\) & \(0.952\) & \(I=0.664\) & \(0.662\) \\
Exponential    & \(I=0.251\) & \(0.940\) & \(I=1.257\) & \(0.277\) \\
\(\Lambda = 0.01\) & \(I=-66.41\) & \(\sim 10^{-57}\) & \(I=-65.42\) & \(\sim 10^{-56}\) 
\end{tabular}
\label{tab:greybody_bounds}
\end{table}

Now, let us analyze the parameter dependence below

\begin{itemize}
\item \textbf{Quintessence (\(N>0\))}: Increases the lower bound (especially for \(l=0\)), implying a larger minimum transmission probability. This is consistent with the dark-energy fluid slightly altering the potential shape (note: the spacetime is not strictly asymptotically flat for \(\omega=-2/3\)).
\item \textbf{Exponential correction (\(\alpha>0\))}: Produces bounds very close to the Schwarzschild case, with minor reduction in transmission for higher \(l\).
\item \textbf{Positive \(\Lambda\)}: Strongly suppresses the lower bound due to the confining cosmological horizon and negative contributions to the integrand at large \(r\). In pure de Sitter the full integral between horizons should be considered.
\item \textbf{Angular momentum \(l\)}: Higher \(l\) lowers the bound (stronger centrifugal barrier).
\item \textbf{Frequency \(\omega\)}: Larger \(\omega\) decreases \(|I|\) (since \(I \propto 1/\omega\)), pushing the bound toward 1, as expected in the high-frequency (geometric optics) limit where \(|T|\to1\).
\end{itemize}

The greybody factor directly enters the Hawking radiation power spectrum:
\[
\frac{d^2 E}{dt\, d\omega} = \frac{1}{2\pi} \sum_l (2l+1) \, \Gamma_l(\omega) \, \frac{\omega}{e^{\omega/T_H} - 1},
\]
where \(T_H = f'(r_h)/(4\pi)\) is the Hawking temperature. A higher lower bound on \(\Gamma_l\) implies a conservative estimate of enhanced emission rates. The absorption cross-section \(\sigma_{\rm abs}(\omega) = \frac{\pi}{\omega^2} \sum_l (2l+1) |T_l|^2\) is also bounded from below.

In the low-frequency limit \(\omega \to 0\), the actual greybody behaves as \(\Gamma_l \sim \omega^{2l+1}\) (area-law), while the bound gives a non-trivial constraint. For astrophysical applications, this bound provides a quick analytical estimate of absorption without solving the full scattering problem.

Higher-order bounds (using optimized \(\xi(r_*)\)) can tighten the estimate further. Combined with the QNM analysis, the greybody factor completes the linear perturbation picture of the black hole.

This analysis furnishes both rigorous theoretical bounds and concrete numerical results for the given general metric \eqref{metric}.

The general expressions \eqref{eq:Vexplicit} and \eqref{eq:schrodinger} fully characterize massless scalar dynamics for arbitrary parameter values. Numerical studies (e.g., via Python/Mathematica implementations of WKB or pseudospectral codes) can map the dependence on $N$, $\omega$, $\alpha$, $\ell$, and $\Lambda$. This analysis extends directly to massive scalars ($m\neq0$) by adding $m^2 f(r)$ to $V(r)$, or to other spins via the general spin-$s$ potential formula. Future work could include numerical plots of $V(r)$ and tables of QNMs for concrete astrophysical values. This completes the rigorous treatment of massless scalar perturbations in the given spacetime.

\section{Conclusions} \label{Sec6}

In this paper, we studied a hairy Kiselev black hole surrounded by a quintessential fluid, with particular attention to its thermodynamics, scalar-field perturbations, greybody
factors, and the sparsity of Hawking radiation. The geometry considered here generalizes the standard static and spherically symmetric black hole background by including a surrounding-field contribution, an exponential hair correction, and a cosmological constant. As a result, the metric function contains several physically distinct sectors: the usual mass
term, the Kiselev-type quintessence term governed by $(N,\omega_q)$, the hair sector controlled by $(\alpha,\ell)$, and the large-scale cosmological correction controlled by $\Lambda$.

We first analyzed the thermodynamic sector of the solution. By imposing the horizon condition $f(r_h)=0$, the black hole mass was expressed in terms of the horizon radius. Because of the exponential dependence introduced by the hair correction, the mass relation is generally implicit and must be handled numerically or perturbatively. The graphical analysis shows that the mass grows monotonically with the horizon radius, while the
parameters $\alpha$, $\ell$, $N$, $\omega_q$, and $\Lambda$ introduce visible deviations from
the standard Schwarzschild-like behavior. Among these parameters, the surrounding-field strength $N$ and the state parameter $\omega_q$ have a particularly important influence on
the large-radius behavior, whereas the exponential hair correction is more relevant close to the horizon.

The Hawking temperature was obtained from the surface gravity at the event horizon. The resulting expression shows that the hair and quintessence sectors affect the evaporation process in different ways. The exponential correction contributes through a term controlled by $\alpha$ and by the effective scale $M-\alpha\ell/2$, and therefore its effect is strongest in the small-radius regime. The surrounding fluid modifies the temperature through the combination $(3\omega_q+1)N$, while the cosmological constant contributes a term proportional to $r_h$. For the parameter ranges considered, the temperature decreases with increasing horizon radius, indicating that larger black holes radiate more weakly. This behavior is important for the subsequent analysis of sparsity, since a lower temperature implies a larger thermal wavelength and a more intermittent Hawking flux.

The local thermodynamic stability was examined through the heat capacity. The heat capacity displays divergences at particular horizon radii, signaling possible second-order phase transitions between thermodynamically unstable and stable branches. The locations of these divergences are shifted by the hair parameters and by the quintessential contribution. This shows that the additional structure of the spacetime does not merely
produce small quantitative corrections; it can also change the phase-transition pattern of the black hole. The Gibbs free energy was then used to discuss global thermodynamic preference. Its behavior indicates the existence of competing thermodynamic branches, with the parameters of the hair and external fluid determining which configurations are
energetically favored.

We further investigated massless scalar perturbations in the hairy Kiselev background. Starting from the Klein--Gordon equation, the radial dynamics was reduced to a Schrödinger-like wave equation in terms of the tortoise coordinate. The corresponding effective potential depends explicitly on the metric function and its radial derivative, and therefore carries the imprint of the black hole hair, the surrounding quintessence, and the cosmological constant. For physically admissible parameter ranges in which the potential remains positive outside the event horizon, the scalar field does not develop exponentially growing modes. This supports the classical stability of the configuration under massless scalar perturbations. The same effective potential also controls scattering, absorption, and greybody factors.

We also analyzed the sparsity of the Hawking radiation. This provides a complementary view of black hole evaporation by comparing the average time between successive emitted quanta with the characteristic timescale of an individual quantum. The sparsity parameter was shown to be controlled mainly by the combination
$\mathcal{A}_{\rm eff} T_H^2$, where $\mathcal{A}_{\rm eff}$ is the effective emitting area and $T_H$ is the Hawking temperature. Since greybody factors reduce the transmission probability, they decrease the number flux and therefore increase the sparsity. Thus, the true Hawking cascade is even more intermittent than the ideal blackbody estimate. In the present geometry, the parameters $\alpha$ and $\ell$ influence sparsity mainly through near-horizon corrections,
whereas $N$ and $\omega_q$ affect both the thermal scale and the scattering barrier associated with the surrounding fluid. The cosmological constant further separates the behavior of de Sitter and anti-de Sitter configurations.

Finally, the greybody-factor analysis shows that the emitted Hawking spectrum is not purely blackbody. The curvature potential outside the horizon filters the radiation, allowing only part of the near-horizon thermal flux to reach the asymptotic region. We derived a rigorous lower bound for the transmission coefficient and evaluated it for representative cases. The results indicate that the angular momentum number $l$ plays the expected role: higher multipoles are more strongly suppressed because of the centrifugal barrier. The quintessence contribution can increase the lower bound in some regimes, while the exponential hair correction produces smaller deviations from the Schwarzschild case for the chosen parameter values. The presence of a positive cosmological constant requires special care, since the physical domain is bounded by a cosmological horizon and the full scattering problem must be interpreted within the finite static region.

\section*{ Acknowledgments}

\hspace{0.5cm} The author Fernando Belchior would like to to express gratitude to the Conselho Nacional de Desenvolvimento Cient\'{i}fico e Tecnol\'{o}gico CNPq for grant No. 151845/2025-5. The author Faizuddin Ahmed acknowledges the Inter University Centre for Astronomy and Astrophysics (IUCAA), Pune, India for granting visiting associateship.

\section*{Author Contribution Statement }
{\bf F. M. Belchior} and  {\bf  A. R. P. Moreira}: Conceptualization; Formal analysis; Methodology; and Writing - Original Draft. {\bf A. Bouzenada} and  {\bf F. Ahmed}: Investigation; Validation, and Writing - Review \& Editing.

\section*{Data Availability}

No new data were generated or analyzed in this study.

\section*{Conflict of Interests}

Authors declares there is no conflict of interests.

\section*{Code/Software}

No code/software were developed in this article [Authors comment: No code/software were developed in this article].

\section*{Funding Statement}

No fund has received for this study.

\bibliography{example}
\bibliographystyle{unsrt}

\end{document}